\documentclass[%
 reprint,	% True two column version
superscriptaddress,
showpacs,preprintnumbers,
 amsmath,amssymb,
prb,
floatfix,
]{revtex4-1}

\usepackage{graphicx}% Include figure files
\usepackage{dcolumn}% Align table columns on decimal point
\usepackage{bm}% bold math
\begin{document}

\preprint{APS/123-QED}

\title{Thermoelectric properties of finite graphene antidot lattices}

\author{Tue Gunst}
\email{Tue.Gunst@nanotech.dtu.dk}
\affiliation{Department of Micro- and Nanotechnology, Technical University of Denmark, DTU Nanotech}
\author{Troels Markussen}
\affiliation{Department of Physics, Technical University of Denmark, DTU Physics}%
\author{Antti-Pekka Jauho}
\author{Mads Brandbyge}
\affiliation{Department of Micro- and Nanotechnology, Technical University of Denmark, DTU Nanotech}

\date{\today}% It is always \today, today,

\begin{abstract}
We present calculations of the electronic and thermal transport properties of  graphene antidot lattices with a finite length along the transport direction.  The calculations are based on the $\pi$-tight-binding model and the Brenner potential. We show that both electronic and thermal transport properties converge fast toward the bulk limit with increasing length of the lattice: only a few repetitions ($\simeq 6$) of the fundamental unit cell are required to recover the electronic band gap of the infinite lattice as a transport gap for the finite lattice.
We investigate how different antidot shapes and sizes affect the thermoelectric properties. The resulting thermoelectric figure of merit, $ZT$, can exceed $0.25$, and it is highly sensitive to the atomic arrangement of the antidot edges. Specifically, hexagonal holes with pure zigzag edges lead to an order-of-magnitude smaller $ZT$ as compared to pure armchair edges. We explain this behavior as a consequence of the localization of states, which predominantly occurs for zigzag edges, and of an increased splitting of the electronic minibands, which reduces the power factor $S^2 G_e$ ($S$ is the Seebeck coefficient and $G_e$ is the electric conductance).
\end{abstract}
\pacs{85.80.Fi,73.23.Ad,68.70.Lm,68.65.Cd,73.21.Cd,63.22.Rc,72.80.Vp}% PACS, the Physics and Astronomy
%63.22.Rc Phonons in graphene. 72.80.Vp Electronic transport in graphene
\keywords{Thermoelectrics, graphene, antidot lattice, nanomesh, ZT, phonon, transport}%Use showkeys class option if keyword
\maketitle
%\tableofcontents

\section{Introduction}%Hvorfor interessant. Metode: elektron struktur, termiske egenskaber, ZT. Anvendelser. Hypotese.
% Thermoelectrics generelt og Seebeck coefficienten i nanostrukture:
Ideal thermoelectric materials conduct electricity very well while the heat conduction is poor.
Their applications include power generation and refrigeration\cite{saqr_critical_2009,lon_e._bell_cooling_2008}. The optimization of thermoelectric properties has been a topic with wide interest\cite{mahan_best_1996,jeong_best_2011}, and
in particular nanostructured materials for thermoelectrics is a rapidly expanding field of research.
One proposal has been to increase the Seebeck coefficient, $S$, by reducing the dimensionality of the system\cite{kim_influence_2009,pichanusakorn_nanostructured_2010,dresselhaus_new_2007}. Another idea is to utilize the low thermal conductance, together with a sharp resonance in the electronic conductance $G_e$, of molecular junctions\cite{saha_multiterminal_2011,dubi_colloquium:_2011,reddy_thermoelectricity_2007,paulsson_thermoelectric_2003}. Reduction of the thermal conductivity in nanostructured materials may be achieved using nanomesh structures, surface-disorder and -decoration, passivation, or by other means. However, the electronic conductance should ideally not be affected.
Examples of nanostructured thermoelectric materials include passivated Si nanowires\cite{markussen_surface-decorated_2009}, Si antidot lattices\cite{tang_holey_2010,yu_reduction_2010}, chevron-type graphene nanoribbons\cite{chen_thermoelectric_2010}, and connected capped carbon nanotubes\cite{esfarjani_thermoelectric_2006}. Here, we turn the attention towards graphene antidot lattices (GALs), a nanomesh of holes in graphene with promising electronic properties such as a tunable band gap\cite{pedersen_graphene_2008}.

The efficiency in converting temperature gradients into an electric voltage, at an average temperature $T$, is quantified by the dimensionless figure of merit $ZT={S^2G_eT}/\kappa$, where high $ZT$ implies a good thermoelectric. We thus seek a high electronic power factor, $S^2 G_e$,  and minimal thermal conductance $\kappa={\kappa_{ph}+\kappa_e}$ which includes contributions both from phonons and electrons. %(Markussen et al. \cite{markussen_surface-decorated_2009}),(Pedersen et al.\cite{pedersen_graphene_2008})
%Thermoelectric materials with $ZT\approx 1$ have been realized from nanostructered bulk materials, whereas $ZT > 3$ is needed to compete with conventional refrigerators and %generators\cite{vining_inconvenient_2009,majumdar_thermoelectricity_2004}.
Thermoelectric materials with $ZT\approx 1$ have an efficiency in the range of available thermoelectric components based on nanostructered bulk materials, whereas $ZT > 3$ is needed to compete with conventional refrigerators and generators\cite{vining_inconvenient_2009,majumdar_thermoelectricity_2004}.

% Grafen:
Graphene can sustain current densities six orders of
magnitude larger than copper, has a measured record high stiffness, and is foreseen to have numerous applications ranging from nanoelectronics,
spintronics and nanoelectromechanical devices\cite{geim_graphene:_2009}.
%(Thermal conductance is an order of magnitude larger than copper\cite{balandin_superior_2008})
Graphene is furthermore one of the best thermal conductors known\cite{balandin_superior_2008,seol_two-dimensional_2010}. It has been predicted to posses a giant Seebeck coefficient when gated by a sequence of metal electrodes\cite{dragoman_giant_2007}. However, ways to reduce the superior thermal conductivity of graphene are needed if one looks for thermoelectric applications.
Several ways to reduce the thermal conductivity have already been examined, such as interface mismatching between graphene and nanoribbons\cite{huang_simulation_2009,huang_simulation_2010}, the presence of isotopes\cite{hu_tuning_2010,mingo_cluster_2010,zhang_isotope_2010,jiang_isotopic_2010} and point defects\cite{ouyang_theoretical_2009,adamyan_phonons_2011}.
Edge disorder has been predicted theoretically to suppress heat conductance of graphene nanoribbons\cite{li_phonon_2010,savin_suppression_2010} and $ZT$ exceeding 3 has been theoretically predicted for such systems in the diffusive limit\cite{sevincli_enhanced_2010}.

% GALs:% Antidot resultater
Graphene antidot lattices (GALs) have been proposed as a flexible platform for creating a semiconducting material with a band gap, which can be tuned by varying the antidot size, shape, or lattice symmetry\cite{pedersen_graphene_2008,fuerst_electronic_2009,Petersen_ACSNano_2011,Ouyang_ACSNano_2011}. %nanomesh of holes in graphene
GALs can be fabricated by electron beam lithography\cite{eroms_weak_2009,Beg_NanoLett_2011},
by block copolymer lithography\cite{kim_fabrication_2010,bai_graphene_2010} with hole distances down to 5nm, and at a larger scale through nanorod photocatalysis\cite{akhavan_graphene_2010} and anisotropic etching\cite{krauss_raman_2010}. % nanoimprint
To the best of our knowledge, no studies have been reported on the thermal properties of finite GALs.  Apart from their intrinsic scientific interest, these studies are necessary to assess whether the thermal properties can be engineered in a manner similar to the electronic case.  Of course, all realistic devices are of finite length, and the study of size-effects is important for practical purposes.
For completeness, we mention here other related studies that have recently been reported.
%The detailed effect of the regular nano-scale perforation on the thermal properties of a graphene layer is interesting for thermoelectric applications.
A number of studies of electron and/or phonon transport properties of regular defects in ribbons are available, see, e.g., Refs.~[\onlinecite{zhang_band_2010,hu_thermal_2009,hancock_generalized_2010}].
Recently, Lopata \textit{et al.} studied electron transport of infinite GALs\cite{lopata_graphene_2010}. Finally, during the preparation of this manuscript Karamitaheri \textit{et al.} reported a combined study of electron and phonon transport properties based on the band structures of infinite GALs\cite{karamitaheri_investigation_2011} and Tretiakov \textit{et al.} reported results for topological insulators\cite{tretiakov_holey_2011} which share certain key properties (e.g., flat bands) with GALs.
%So far theoretical work on GALs have focused on the properties of infinite systems. However, in realistic systems the length of the device is finite, and it %is therefore interesting to examine the effect of the finite size on the transport properties.

% Mål og motivation:%Can we engineer one of the best thermal conductors known to state of the art thermoelectric applications?
The topic of this paper is thus the electronic and thermal transport properties of finite graphene antidot lattices. The finite GALs are viewed as a part of an integrated graphene-based system, e.g. used as an electrode for molecular conductors\cite{saha_multiterminal_2011,lue_quantum_2008}, see Fig.~\ref{fig:Unitcells}.
In order to shed light on the question to what extent it may be possible to engineer the thermoelectric properties we investigate how different antidot shapes and sizes affect the thermoelectric properties. Interestingly, even though the base material - graphene - is an outstanding thermal conductor, we find that the resulting thermoelectric figure of merit, $ZT$, can exceed $0.25$. However, $ZT$ is highly sensitive to the atomic arrangement at the edge of the etched holes, partly due to electronic quasi-localized edge-states. As we shall show below, this favors antidots with armchair-type edges for thermoelectrics.

% Organisering:
The paper is organized as follows. In section \ref{System} we introduce the systems and outline the theoretical and numerical methods used. In section \ref{Electron} we present our results for the electronic transport properties of GALs. Especially, we discuss the convergence with number of repetitions of the basic unit cell, and also consider the localization of electronic states at zigzag edges. In Section \ref{Phonon} we examine the influence of the perforation and geometrical effects on the thermal transport properties. This leads to the analysis of thermoelectric properties in section \ref{ZT}. The results are summarized and discussed in section \ref{Conclusion}.
%topological modification from\cite{balandin_superior_2008,seol_two-dimensional_2010}

\section{Systems and Methods \label{System}}
\begin{figure}[htbp]
\centering
%{\includegraphics[width=0.3\paperwidth]{FiguresInitial/Unitcells.png}}
%{\includegraphics[width=0.25\paperwidth]{FiguresInitial/UnitcellsV2.png}}
%{\includegraphics[width=0.25\paperwidth]{FiguresInitial/SystemFigNewV2}}
%{\includegraphics[width=0.4\paperwidth]{FiguresInitial/SystemFigNewZoomCrop.png}}
%{\includegraphics[width=0.4\paperwidth]{SystemFigNewZoomHexaLeadsV3}}
%{\includegraphics[width=0.42\paperwidth]{SystemFigNewZoomHexaLeadsV5}}
{\includegraphics[width=0.42\paperwidth]{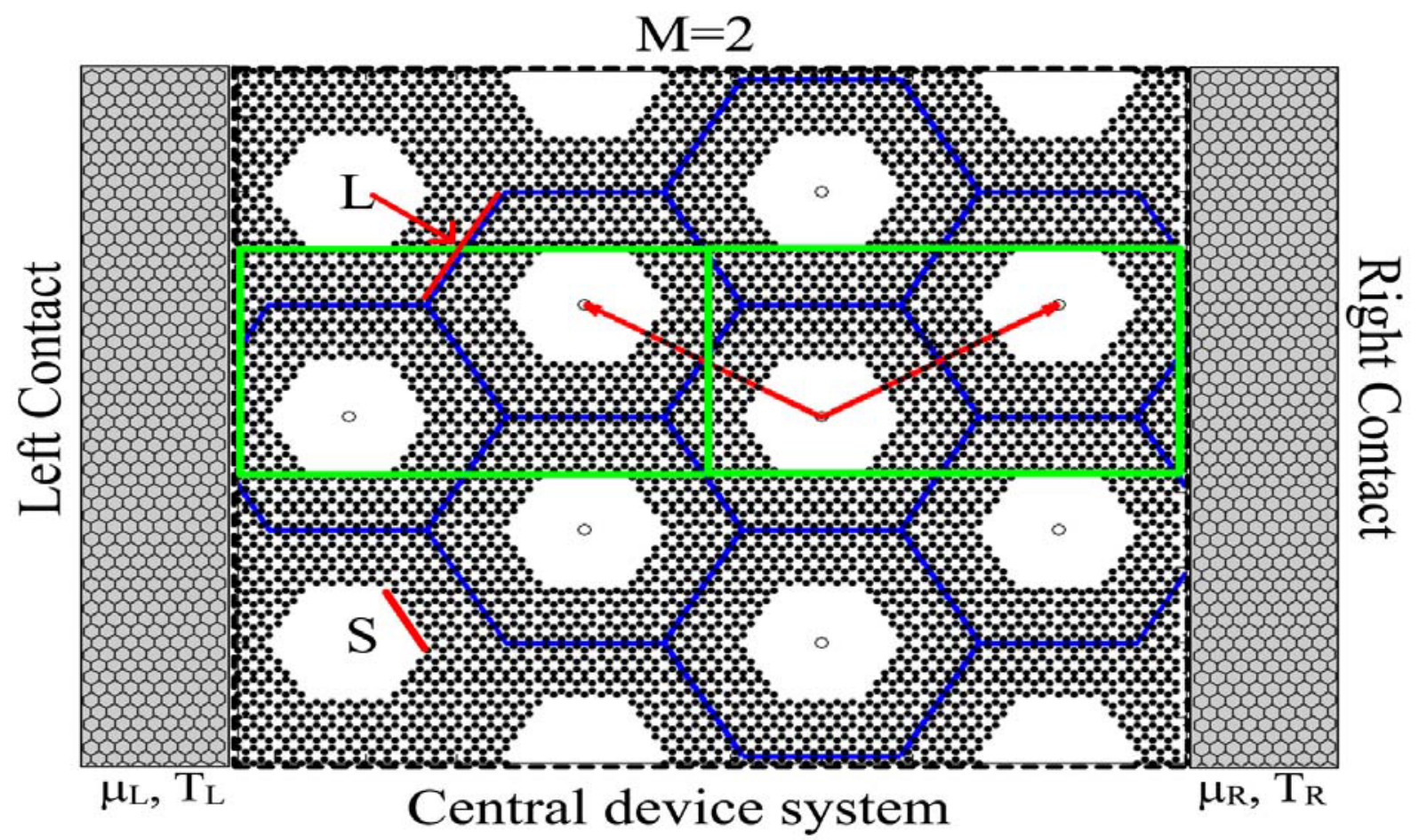}}
\caption{(Color online) System setup and the computational rectangular unit cell (green rectangle). Two graphene leads are connected by the finite GAL. The depicted system is a \{10,5zz\} GAL with a length of 2 ($M=2$) corresponding to 4 holes in the direction of transport.}
\label{fig:Unitcells} %  in a triangular lattice
\end{figure}

% 2D Graphene band structure, square unitcell, symmetry points. Transport unitcell.
%Metode: elektron struktur->pi model&HarrisonScaling&, termiske egenskaber->Brenner&Harmonisk&Impulbevarelse,Both->AtomisticGF&Reuse of single cell matrix to converge in length, ZT-> eqn's.
%System-> square unitcell&GrapheneDispersion&SystemWith3DifferentHoles...Notation.
Throughout in this paper we  focus  on triangular graphene antidot lattices: these systems are known to lead to a gap in the electronic spectrum\cite{Petersen_ACSNano_2011,Ouyang_ACSNano_2011} which is essential for the present purposes. Due to the high lattice symmetry the number of independent lattice parameters is small, and furthermore, these systems are the most thoroughly studied, both theoretically and experimentally. Recent experiments have illustrated that hexagonal antidots may stabilize with pure zigzag and armchair edge chirality\cite{krauss_raman_2010}. The antidot diameter, shape, position and the ratio of removed atoms to unit cell size are all important parameters which we examine to get a full picture of the electronic, thermal and thermoelectric properties of GALs. Another important variable is the length of the region exposed to the nanoperforation. The systems studied here consist of an antidot lattice of finite length connected to two pristine graphene leads (Fig.~\ref{fig:Unitcells}), and the infinite direction perpendicular to the transport direction is treated using periodic boundary conditions and corresponding $k$-point sampling.
%and an important issue to address is how fast the properties of finite GAL converge to that of the infinite antidot lattice as the %length is increased.

We use the nomenclature introduced in Ref.~\onlinecite{pedersen_graphene_2008} and specify the GAL by \{$L$$s_1$,$S$$s_2$\}, where the $L$ is the length of the side of a hexagonal Wigner-Seitz cell, and $S$ is the side length of the antidot(see Fig.~\ref{fig:Unitcells}), both in units of the lattice constant $\sqrt{3} a_0$, with $a_0=1.42\,$\AA \, being the carbon-carbon distance. The label $s_2=\{{\rm zz},{\rm arm},{\rm cir}\}$ indicates whether the hole has zigzag or armchair edges, or if the hole is circular resulting in mixed armchair and zigzag edges. In principle, one could also consider different sheet orientations compared to the transport direction; we do not present a systematic study here, and fix the underlying graphene sheet as armchair ($s_1={\rm{arm}}$, as in Fig.~\ref{fig:Unitcells}). We have tested a selection of 'zigzag sheets' and did not find any qualitative difference with respect to thermoelectric properties.
As an example \{10,5zz\} is a $L=10$ antidot lattice with transport direction perpendicular to the armchair direction\cite{Note1} and with a hexagonal hole with same orientation as the lattice hexagons resulting in zigzag edges and a side length of $S=5$ (see Fig.~\ref{fig:Unitcells}). Armchair edges are obtained if the hexagonal holes are rotated opposite to the lattice hexagons (see also Fig.~\ref{fig:Relaxations} below).%{fig:TeBands}

\subsection{Method}
Both electronic and phonon transport properties are calculated from  atomistically determined energy-dependent transmission functions, $\mathcal{T}_e$ and $\mathcal{T}_{ph}$, as described below, and using these in a Landauer-type formula.
For a spin degenerate electronic system the Landauer formula reads,
\begin{eqnarray}
I_e = \frac{2e}{\hbar}\int \frac{dE}{2\pi}\mathcal{T}_e(E) [n_F(E,\mu_L)-n_F(E,\mu_R)]\,,\label{eqn:Landauer}
\end{eqnarray}
where $n_F(E,\mu_{L/R})$ is the Fermi-Dirac distribution at the chemical potential of the left/right lead.
We will employ this in the linear-response limit, and consider variations with changes in the chemical 
potential, e.g. by doping or gating of the graphene system.
The following integrals can be evaluated from the electronic transmission,
\begin{equation}
L_n(\mu) = \frac{2}{\hbar}\int\frac{dE}{2\pi}(E-\mu)^n\mathcal{T}_e(E)\left(-\frac{\partial n_F}{\partial E}\right).\label{eqn:Ln}
\end{equation}
They relate the electronic current and the electron heat current, $I_{Q}$, in the linear response regime:
\begin{equation}
\left(\begin{array}{cc}
\frac{\Delta I_e}{e}\\
\Delta I_{Q}\end{array}\right)
=	
\left(\begin{array}{cc}
L_0 & L_1\\L_1 & L_2
\end{array}\right)
\left(\begin{array}{cc}
\Delta \mu\\ \frac{\Delta T}{T}
\end{array}\right)\,,\label{eqn:CouplingMatrix}
\end{equation}
where $\Delta\mu=\mu_L-\mu_R$ and $\Delta T=T_L-T_R$.
%This describes that a small perturbation in temperature difference between the leads induces a voltage difference and vice versa.
From these integrals several physical properties follow\cite{sivan_multichannel_1986};
the electrical conductance $G_e(\mu)=\frac{\partial I}{\partial V}=e^2L_0$, the electron thermal conductance $\mathcal{\kappa}_e(\mu) = \left[L_2-\frac{L_1^2}{L_0}\right]/T$, and the Seebeck coefficient $S(\mu) = \frac{\Delta V}{\Delta T}|_{I_e=0} = \frac{L_1}{eL_0T}$.

For phonons the Landauer formula takes an analogous form,
\begin{eqnarray}
I_{ph} = \int_{0}^{\infty}d\omega \frac{\hbar \omega}{2\pi} \mathcal{T}_{ph}(\omega) \left[n_B(\omega,T_L)-n_B(\omega,T_R)\right]\,,
\end{eqnarray}
where $n_B(\omega)$ is the Bose distribution function.
Again we use it in linear response and consider the thermal conductance from phonons given by,
\begin{eqnarray}
\kappa_{ph} = \int_{0}^{\infty}d\omega \frac{\left(\hbar \omega\right)^2}{2\pi k_B T^2} \mathcal{T}_{ph}(\omega) \frac{e^{\frac{\hbar \omega}{k_B T}}}{(e^{\frac{\hbar \omega}{k_B T}}-1)^2}\,.
\end{eqnarray}

Both transmission functions, $\mathcal{T}_e$ and $\mathcal{T}_{ph}$ are obtained using a recursive Green's function method (see Ref.~\onlinecite{markussen_electronic_2006} and the references cited therein) with self-energies representing the semi-infinite perfect graphene electrodes. The self-energies, $\mathbf{\Sigma}_{L,R}$, are iteratively constructed from the semi-infinite graphene left $(L)$ and right $(R)$ leads.
The calculation of both electron and phonon $k$-averaged Landauer transmissions together with the thermoelectric properties are performed by an atomistic Green's function method\cite{markussen_electron_2009,esfarjani_thermoelectric_2006},
\begin{eqnarray}
\mathcal{T}_e(E) &=& \frac{1}{N_{k}}\sum_{i=1}^{N_k}\textrm{Tr}\left[\mathbf{G}^r_{D}(E,k_i)\mathbf{\Gamma}_{R}(E,k_i)\right.\nonumber\\ &\quad&\times\left.\mathbf{G}^a_{D}(E,k_i) \mathbf{\Gamma}_{L}(E,k_i)\right].
\end{eqnarray}
Here the retarded Green's function  $\mathbf{G}^r_D(E,k)$, is obtained from the Hamiltonian, $\mathbf{H}$, $\mathbf{G}^r_{D}=[E\mathbf{I} -\mathbf{H}-\mathbf{\Sigma}^r_{L}-\mathbf{\Sigma}^r_{R}]^{-1}$, and the broadening matrices due to the electrode coupling are defined as $\mathbf{\Gamma}_{L,R}=i[\mathbf{\Sigma}^r_{L,R}-\mathbf{\Sigma}^a_{L,R}]$. The parameter $N_{k}$ gives the number of sampled $k$-points\cite{Note2}. Similar equations hold for the phonon transmission: the Hamiltonian is replaced with the dynamical matrix $\mathbf{H}\rightarrow \mathbf{K}$, and the energy is replaced with $E\mathbf{I}\rightarrow \omega^2\mathbf{M}$, $\omega$ being the frequency and $\mathbf{M}$ is the diagonal mass matrix.
We first perform a structural relaxation and then calculate the Hamiltonian/dynamical matrix for three unitcells ($M$=3, 6 holes) between the pristine graphene leads. The elements corresponding to the center cell (2 holes) are then subsequently repeated to increase the length of the GAL.

The electronic system is modeled by a nearest-neighbor $\pi$-model ($V_{pp\pi}=2.7\,$eV) together with the Harrison scaling law to take into account the changes in the hopping matrix element due to the edge relaxation\cite{Harrison_book}.
Based on the same method Guinea and co-workers\cite{,guinea_energy_2010} have shown how strain in graphene can lead to a pseudo-magnetic field affecting the electronic properties.
%This modulation of the hopping elements have recently been applied do obtain large pseudo-magnetic fields, and a resulting band %gap, in graphene from an applied strain field\cite{,guinea_energy_2010}.
We find that the modulation of the hopping elements is of minor importance for the present applications. In order to examine the effect of passivation we have performed calculations of the band structures with a model including two $d$-orbitals for each C-atom and an explicit model for the carbon-hydrogen interaction\cite{boykin_accurate_2011}. The qualitative features of the band diagram, and the edge states discussed below, depend surprisingly little on the presence of hydrogen passivation. However, the effect of passivation might be more important in antidot lattices with localized zero energy states such as in the triangular antidots considered in Ref.~\onlinecite{vanevic_character_2009}.

\begin{figure}[htbp]
\centering
%{\includegraphics[width=0.3\paperwidth]{FiguresInitial/Unitcells.png}}%SystemBondlengthsArm10.png}}{\includegraphics[width=0.4\paperwidth]{FiguresInitial/RelaxationsArmZzCir.png}}
{\includegraphics[width=0.35\paperwidth]{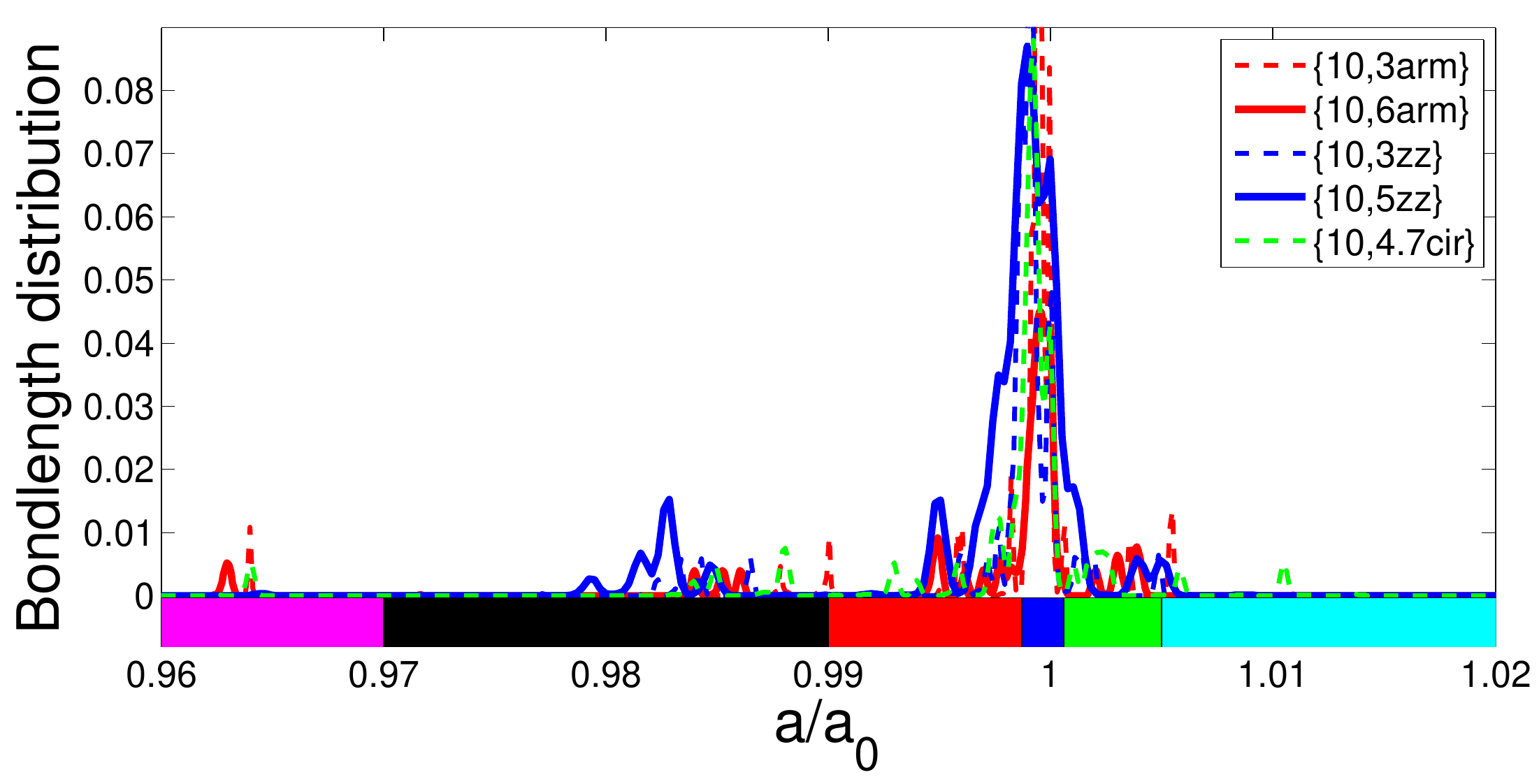}}\\
%{\includegraphics[width=0.35\paperwidth]{SystemBondlengthsM=3ZoomArm5}}
%{\includegraphics[width=0.35\paperwidth]{SystemBondlengthsM=3ZoomArm10}}
%{\includegraphics[width=0.35\paperwidth]{SystemBondlengthsM=3ZoomZz5}}
%{\includegraphics[width=0.35\paperwidth]{SystemBondlengthsM=3ZoomZz9}}
%{\includegraphics[width=0.35\paperwidth]{SystemBondlengthsM=3ZoomCir8}}
{\includegraphics[width=0.38\paperwidth]{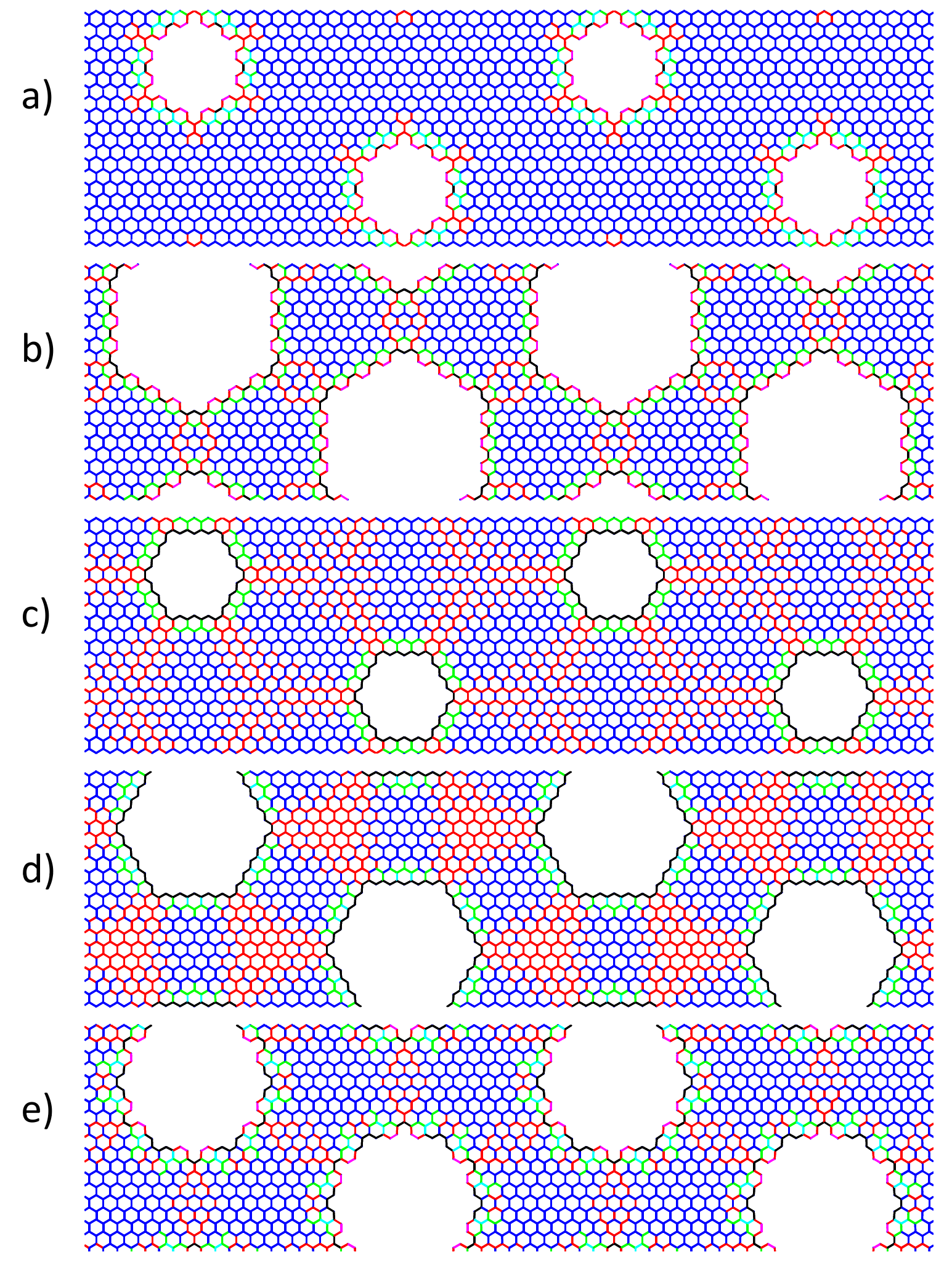}}
\caption{(Color online) Change of bond lengths due to the relaxation of the graphene antidots. Top: Coloring scheme for the bond lengths. The distribution of bond lengths after relaxation is given for five different lattices. From top to bottom the considered lattices have a) small armchair holes \{10,3arm\}, b) large armchair holes \{10,6arm\}, c) small zigzag holes \{10,3zz\}, d) large zigzag holes \{10,5zz\} and e) mixed edges \{10,4.7cir\}. When the hole size is increased the spectrum broaden and peaks occur at different bond lengths. The peaks occur at different positions characteristic of different hole types. The three bonds in an armchair are characterized by a red-pink-red color sequence.}
\label{fig:Relaxations}
\end{figure}

The dynamical matrix is computed using the empirical Brenner interatomic potential\cite{brenner_empirical_1990}. This is done for the system cell by the finite difference approach after a structure relaxation performed by the 'General Utility Lattice Program' (GULP)\cite{gale_general_2003}. Momentum conservation is important for low frequency modes and it is imposed after the finite difference calculation by adjusting the diagonal elements of the dynamical matrix.
% fig med farve fordeling
A few representative results of the relaxation are shown in Fig.~\ref{fig:Relaxations}.
We see that the changes in the bond lengths compared to $a_0$ are below 4\%.
Compression of bonds at the edge is followed by a region with elongation of bonds and the relaxation is confined in small regions in space. Matching of edge relaxations can result in longer-ranged relaxations (small compressions) emanating from the corners. This occurs mainly for zigzag edges (Fig.~\ref{fig:Relaxations}).

The results presented below are based on a set of electron and phonon transport simulations of 20 configurations with varying hole size and lattice parameters. In addition to this set the electronic structure of another 27 systems was studied to examine on the formation of edge states. Finally, a number of systems has been studied with either the zigzag transport direction or a rectangular lattice of holes. No qualitative differences in thermoelectric properties were found for these systems and we shall not present these results here.

\section{Electron transport\label{Electron}}
\begin{figure}[htbp] % Fine now (Change to one that goes beoynd M=6,\{10,7zz\} a possibility).
\centering
{\includegraphics[width=0.4\paperwidth]{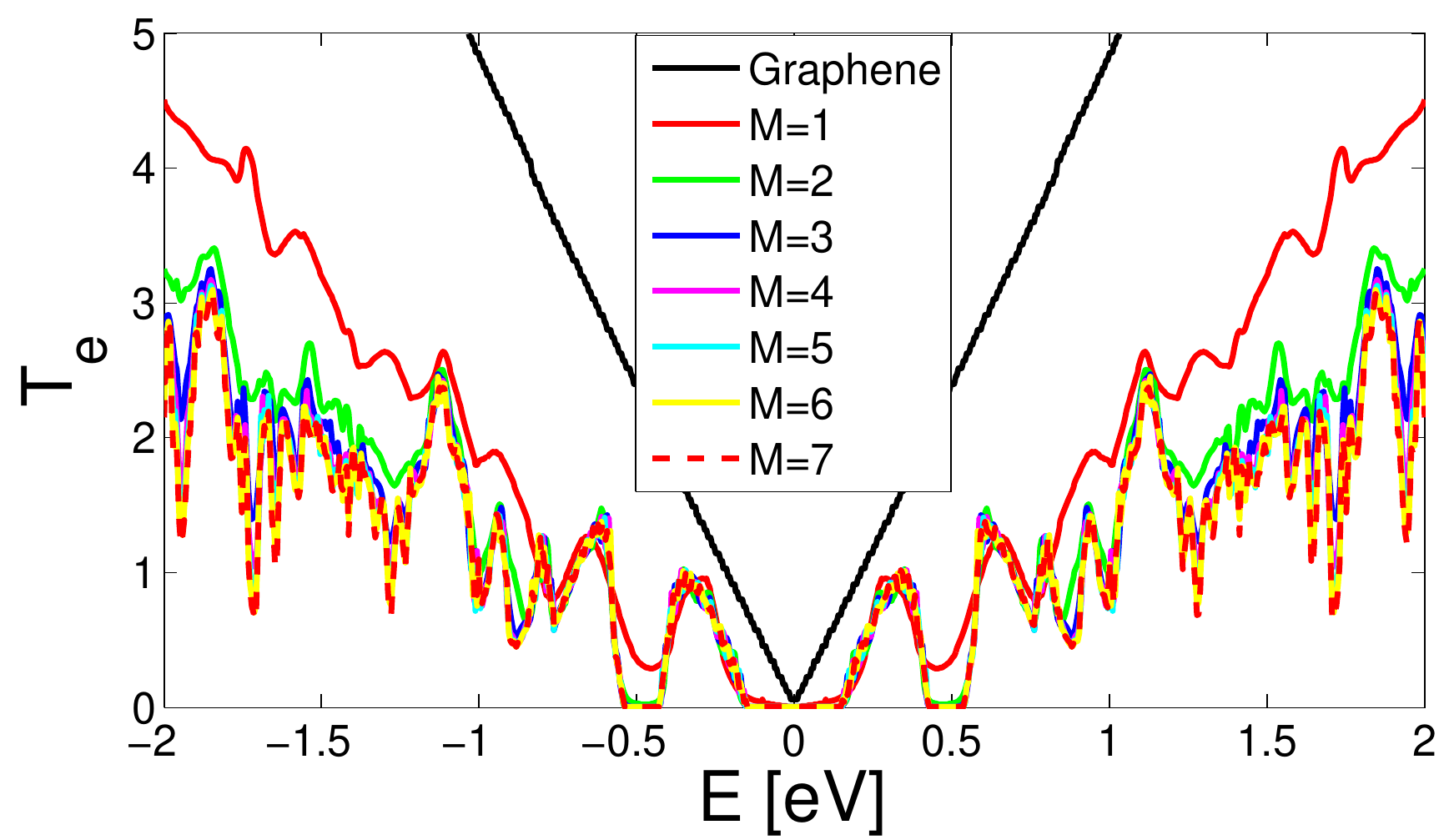}}\\%0.3
%{\includegraphics[width=0.19\paperwidth]{FiguresInitial/EelctronicConductanceCombi.png}}
%{\includegraphics[width=0.3\paperwidth]{FiguresInitial/TransmissionElectronsZoom1.png}}
%{\includegraphics[width=0.19\paperwidth]{FiguresInitial/TransmissionElectronsZoom2.png}}
%{\includegraphics[width=0.4\paperwidth]{FiguresInitial/TransmissionElectronsZoom3New.png}}%0.3
{\includegraphics[width=0.19\paperwidth]{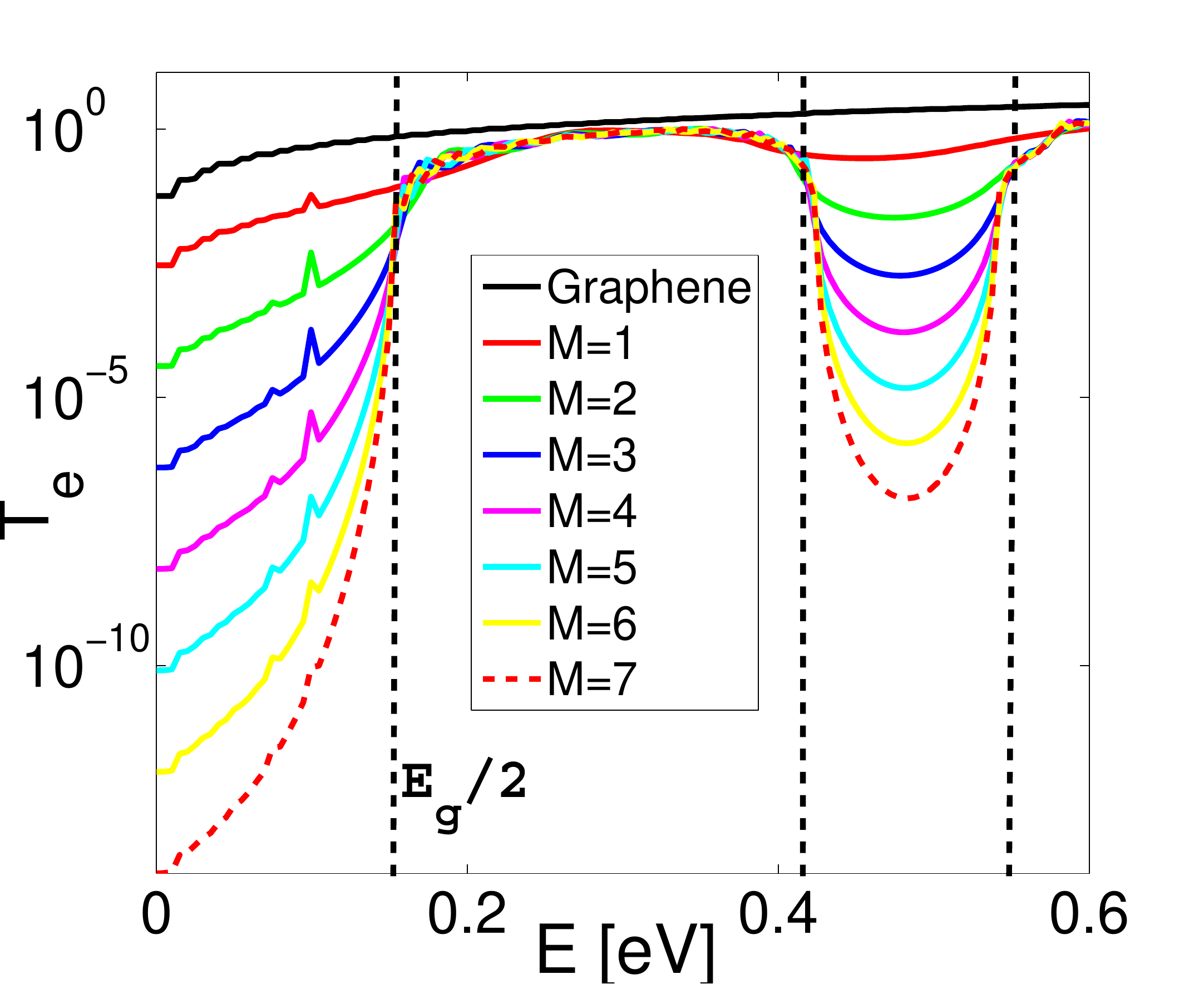}}%0.3
{\includegraphics[width=0.19\paperwidth]{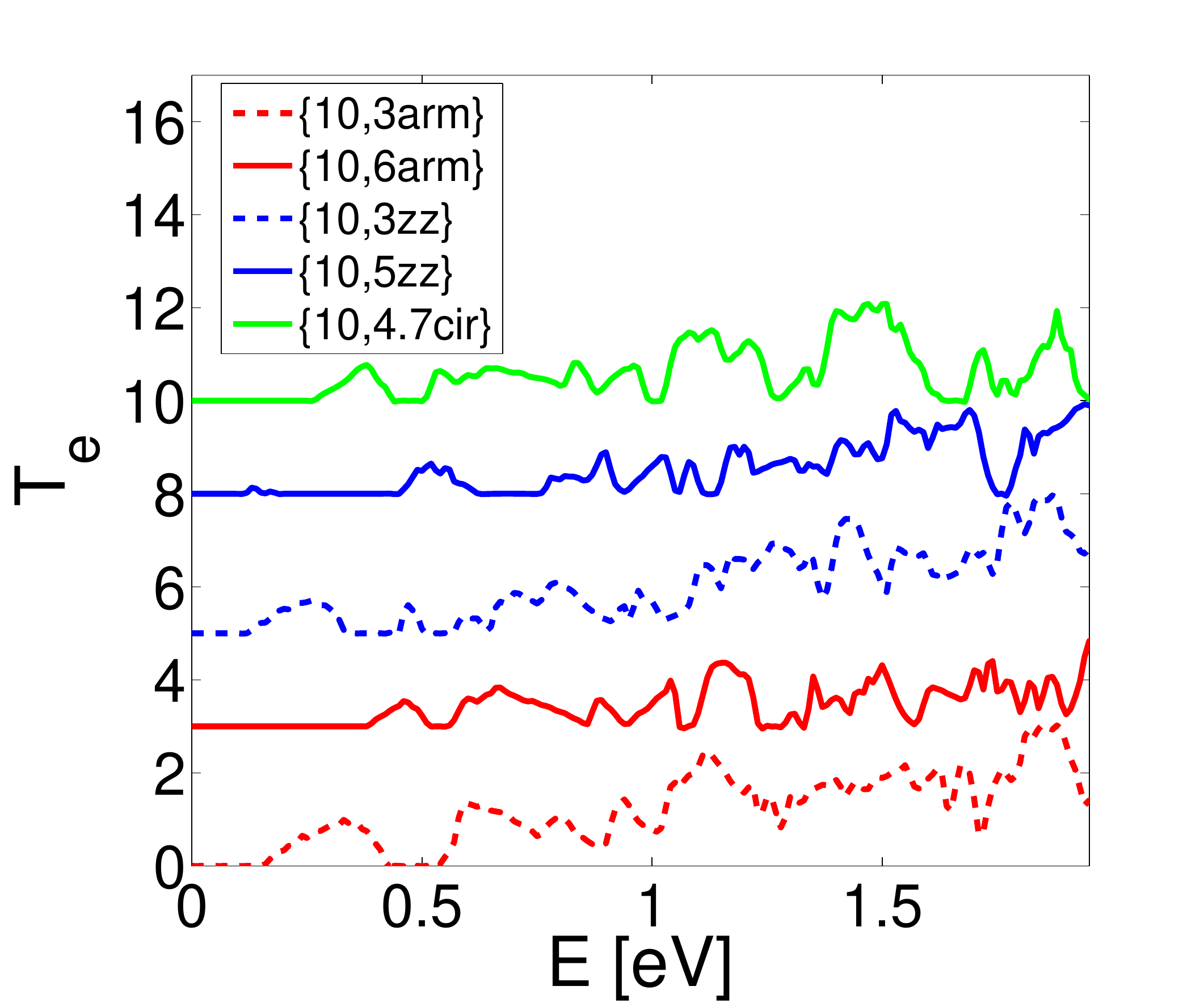}}%0.3
\caption{(Color online) Top: Convergence with length ($M$, the number of unit cells with two holes along the device) of transmission for a \{10,3arm\} antidot lattice. Bottom left: Zoom at the band gap for the \{10,3arm\} GAL. The leftmost vertical dashed line marks the value of the band gap obtained from the band structure of an infinite GAL. Bottom right: Transmissions close to the Fermi level for the selected GALs shown in Fig.~\ref{fig:Relaxations}. The transmission curves have been shifted by an integer to ease the comparison.}
\label{fig:TransConvergence}
\end{figure}

\begin{figure}[htbp]
\centering
%{\includegraphics[width=0.18\paperwidth]{FiguresInitial/EgDependenceV3.png}}
%{\includegraphics[width=0.18\paperwidth]{FiguresInitial/EgDependenceSQRV3.png}}
{\includegraphics[width=0.4\paperwidth]{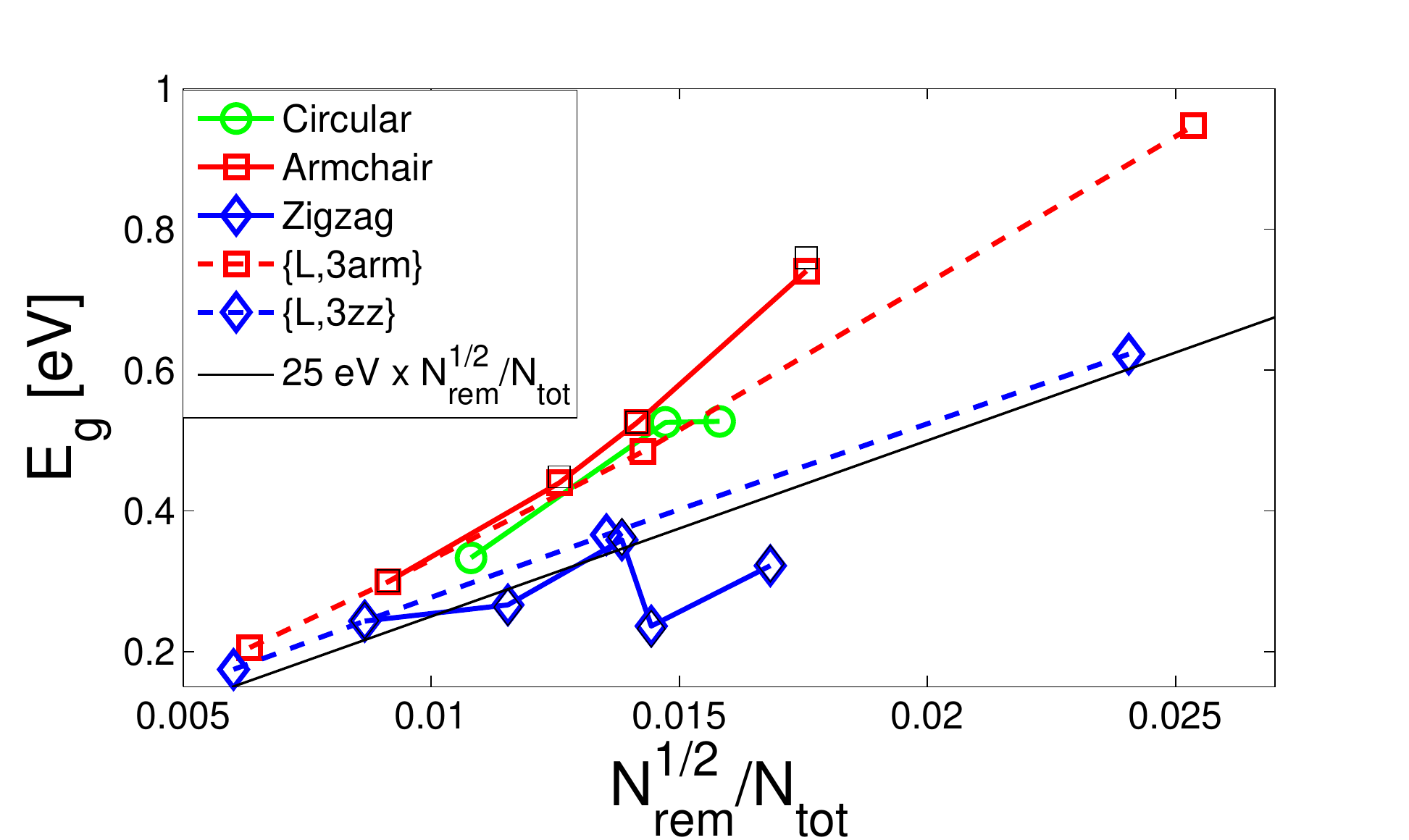}}%0.3
\caption{(Color online) Scaling of the electronic band gap $E_g$ with increasing ratio of removed atoms compared to the simple scaling law estimated for circular holes\cite{pedersen_graphene_2008}. The systems considered are \{10,$S$zz\} with $S=3,4,4.5,5,5.5$, \{10,$S$arm\} with $S=3,4.5,5,6$, and \{10,$S$cir\} with $S=3.5,4.7,5$. Furthermore we include two sets, \{$L$,3arm\} and \{$L$,3zz\} for $L=6,8,10,12$, with fixed hole geometry.}
\label{fig:Egscaling}
\end{figure}

In this section we consider the electronic properties of finite GALs. The result for a series of unit cell repetitions ($M$) is shown in Fig.~\ref{fig:TransConvergence}.
As can be seen the electronic transmission $\mathcal{T}_e$ converges fast toward a length independent result. The behavior of the transmission function can directly be traced back to the band structure of the infinite GAL if one defines a transport band gap as the energy range where the transmission is below a certain small value. The band gap is in general found to converge to that of the infinite antidot lattice found from the band structure and the system behaves 'bulk-like' after only six to seven unit cell repetitions.
Thus the transport band gap can be determined from a calculation of the dispersion on an infinite GAL using a primitive unit cell due to the fast convergence property illustrated in this section. The converged values of the band gap are given in Fig.~\ref{fig:Egscaling}.

To access the effect of relaxation on the electronic structure we have plotted the obtained band gaps neglecting the modulation of the hopping elements for the results with varying hole size for the armchair holes (black squares) and zigzag holes (black diamonds). The relaxation is found not to play a qualitative role in the equilibrium electronic properties of GALs within this model. The convergence is independent of the lattice parameters, and in all simulations presented hereafter between 8 and 10 unit cells are used.

% Short version:
Besides the band gap we observe that it is possible to approximate the transmission versus energy as linear curves corresponding to a simple reduction of the pristine transmission, $\mathcal{T}_0\propto |E|$ (see example in the transmission plots in Fig.~\ref{fig:TeBands}). We have calculated envelope lines obtained from a scaling of the pristine transmission with the width of the constriction $\mathcal{T}_{\rm{eff}}=R_{\rm{eff}}\, \mathcal{T}_{0}$, where $\mathcal{T}_{0}$ is the transmission of pristine graphene. The reduction factor, $R_{\rm{eff}}$, describes the amount of pristine transmission that survives the lattice perturbation in terms of a regular perforation. The actual reduction factor is estimated as the average reduction found at each energy point. We find that the electronic transmission is reduced more than what would be expected from the effective width reduction $R_{\rm{eff}}={W}/{W_0}$. Here $W$ is the minimal width along the device and $W_0$ is the width of the pristine graphene sheet. For the systems considered in Fig.~\ref{fig:Relaxations} the hole dimension is varying between 1.2\,nm and 2.6\,nm giving an effective width reduction between 71\% and 26\%. The actual reduction factor is descreasing linearly with hole dimension from 24\% to 5\%. Therefore, only a minor part of the average transmission reduction can be ascribed to the narrowing of the conducting plane. The present model does not take special account of the band gap opening. One could instead ask if the peak transmission is limited by the effective width. The peak transmission reduction factor is found to be decreasing from 65\% to 21\%, and fits the effective width reduction very well for small holes. As the hole size increase the effective width is overestimated due to the triangular lattice structure of the perforation, and the reduction factor approaches the averaged value.

\subsection{Localization at zigzag edges\label{Sect:Localized}}
\begin{figure*}[htbp]
\centering
{\includegraphics[width=0.65\paperwidth]{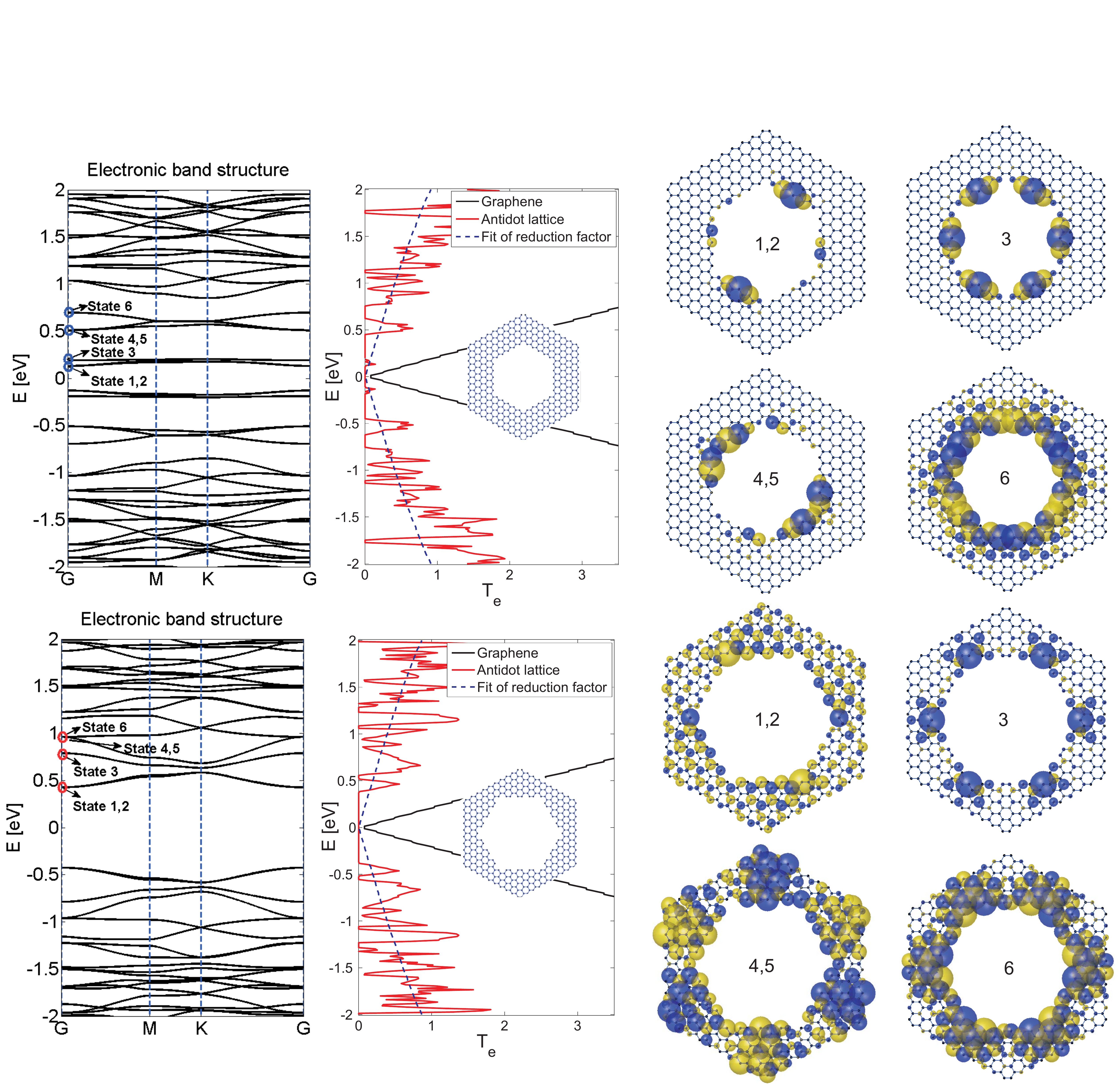}}
%{\includegraphics[width=0.18\paperwidth]{FiguresInitial/ElectronicBandstructureL=10S=3.png}}3,2
%{\includegraphics[width=0.18\paperwidth]{FiguresInitial/ElectronicBandstructureL=10S=5V2.png}}
%{\includegraphics[width=0.175\paperwidth]{FiguresInitial/TransmissionElectronsFinalShiftedZzV5.png}}
%{\includegraphics[width=0.18\paperwidth]{FiguresInitial/ElectronicBandstructureL=10S=6V2.png}}
%{\includegraphics[width=0.175\paperwidth]{FiguresInitial/TransmissionElectronsFinalShiftedArmV4.png}}
\caption{(Color online) Left: Band structure for \{10arm,5zz\} (top) and \{10arm,6arm\} (bottom) antidot lattices respectively. Middle: Corresponding electronic transmission around the Fermi level. The energies of the states illustrated and compared further have been marked with circles. Right: Eigenstates 1, 3, 4 and 6 at the $\Gamma$-point with energies as marked in the corresponding band structures. The eigenstates of the \{10,5zz\} antidot lattice (top) are very localized at the edges. The eigenstates of the \{10,6arm\} antidot lattice (bottom) are less localized at the edges, but resemble corner states. A phase of zero and $\pi$ is colored blue and yellow respectively.}
\label{fig:TeBands}%2D route: $G=0$, $A =(\frac{\pi}{Lx},0,0)$, $B=(0,\frac{\pi}{Ly},0)$, $C=(\frac{\pi}{2\,Lx},\frac{\pi}{2\,Ly},0)$.
\end{figure*}

There is an important difference between holes of different atomic arrangements at the edges.
In Fig.~\ref{fig:TeBands} we compare the electronic transmission and the band structures for two large holes with zigzag and armchair edges, respectively. The figure illustrates how the transmission can be directly traced back to the band structure of the GAL. It is furthermore seen how the structure with zigzag edges leads to an additional splitting into flat minibands around the Fermi level. This feature can be understood in terms of localized states due to a local excess of atoms of one of the two sublattices in the graphene bipartite lattice\cite{vanevic_character_2009}.
The local imbalance of A and B type atoms at the edges leads to the corresponding number of defect states.
In hexagonal holes with zigzag edges each side consists of a segment of either type A or B atoms. The hybridization will be small between these defect states which are partially separated in space. As illustrated at the rightmost of Fig.~\ref{fig:TeBands}, the flat minibands are highly localized at the edges.
In the case of hexagonal holes with armchair edges each side consists of an alternating sequence of A and B atoms. Therefore, these defect states hybridize more resulting in a larger shift from the Fermi level and a reduced flatness of the bands. As can be seen from Fig.~\ref{fig:TeBands} the first bands with minimal dispersion are mainly localized in the small zigzag corner region between two AB sequences.

It is possible to quantify the degree of localization from the weight of the eigenstate at each atom.
The localization factor, for a given eigenstate in the site basis $\psi_n=[u_1,u_2,...,u_N]$, is here defined as,\cite{visscher_localization_1972,kaplan_structure_2001}
\begin{eqnarray}
L_{f}(\psi_n)=\frac{\sum_{i=1}^N |u_i|^4}{\left(\sum_{i=1}^N |u_i|^2\right)^2}\,.
\end{eqnarray}
This factor equals $1/N$ when the state is fully delocalized and all weights $u_i$ have the same value. In the case of a state localized at a single site it gives 1. The inverse localization factor gives a measure of the number of sites that contribute to a given state.

%The $L=10$ curve is for the states illustrated.
\begin{figure}[htbp]
\centering
%{\includegraphics[width=0.18\paperwidth]{AntidotHexagonalBands/EdgeLocalizationLdependenceDir=armchairMode=polygonHoletype=zigzagedgeL=10Holedim=9/LocalizationLDependenceV2.png}}
%{\includegraphics[width=0.4\paperwidth]{AntidotHexagonalBands/EdgeLocalizationLdependenceDir=armchairMode=polygonHoletype=zigzagedgeL=10Holedim=9/LocalizationLDependenceStateV2.png}}
%{\includegraphics[width=0.18\paperwidth]{AntidotHexagonalBands/EdgeLocalizationLdependenceDir=armchairMode=polygonHoletype=armchairedgeL=10Holedim=10/LocalizationLDependenceV2.png}}
%{\includegraphics[width=0.4\paperwidth]{AntidotHexagonalBands/EdgeLocalizationLdependenceDir=armchairMode=polygonHoletype=armchairedgeL=10Holedim=10/LocalizationLDependenceStateV2.png}}%0.3
{\includegraphics[width=0.4\paperwidth]{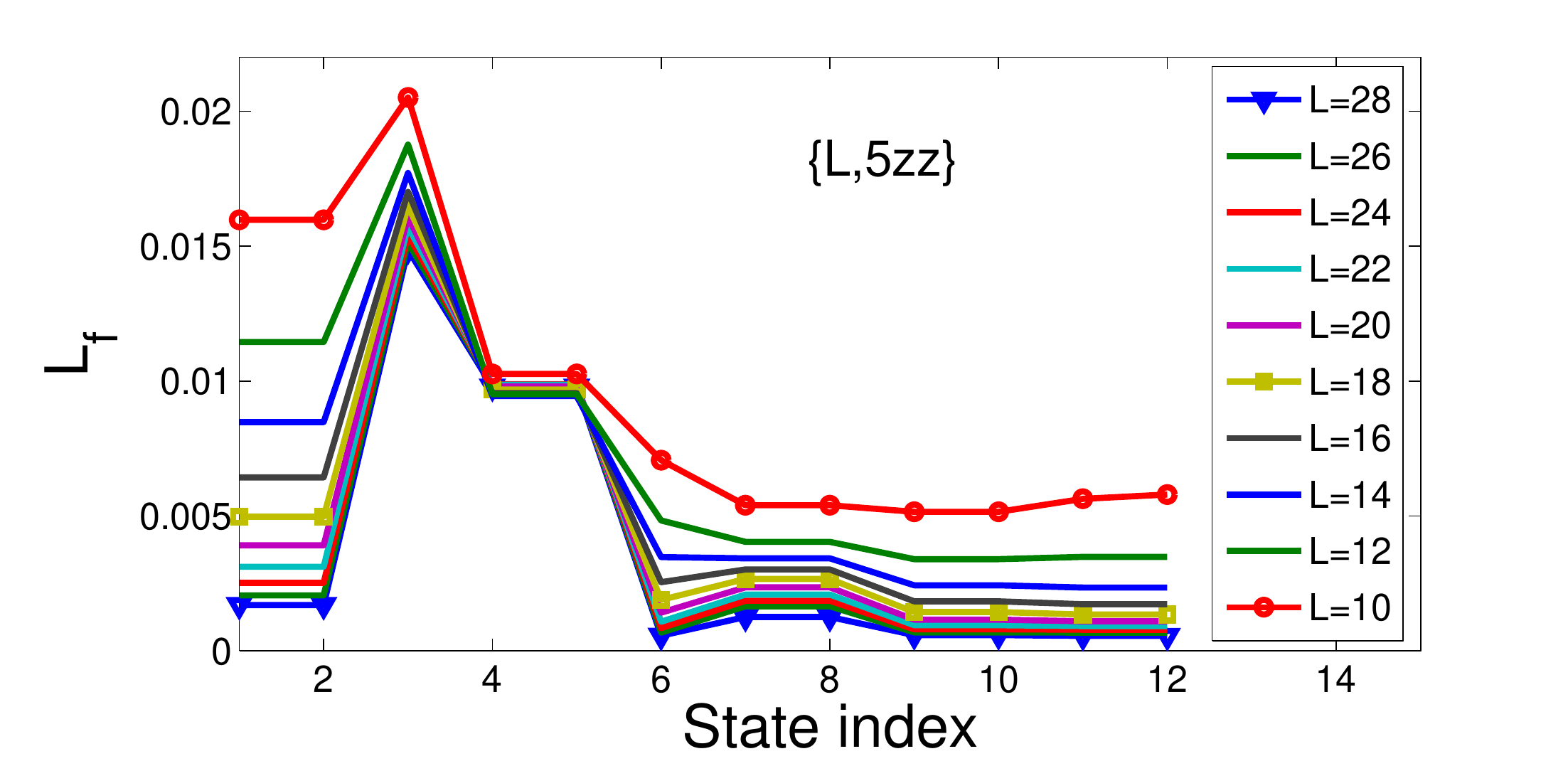}}
{\includegraphics[width=0.4\paperwidth]{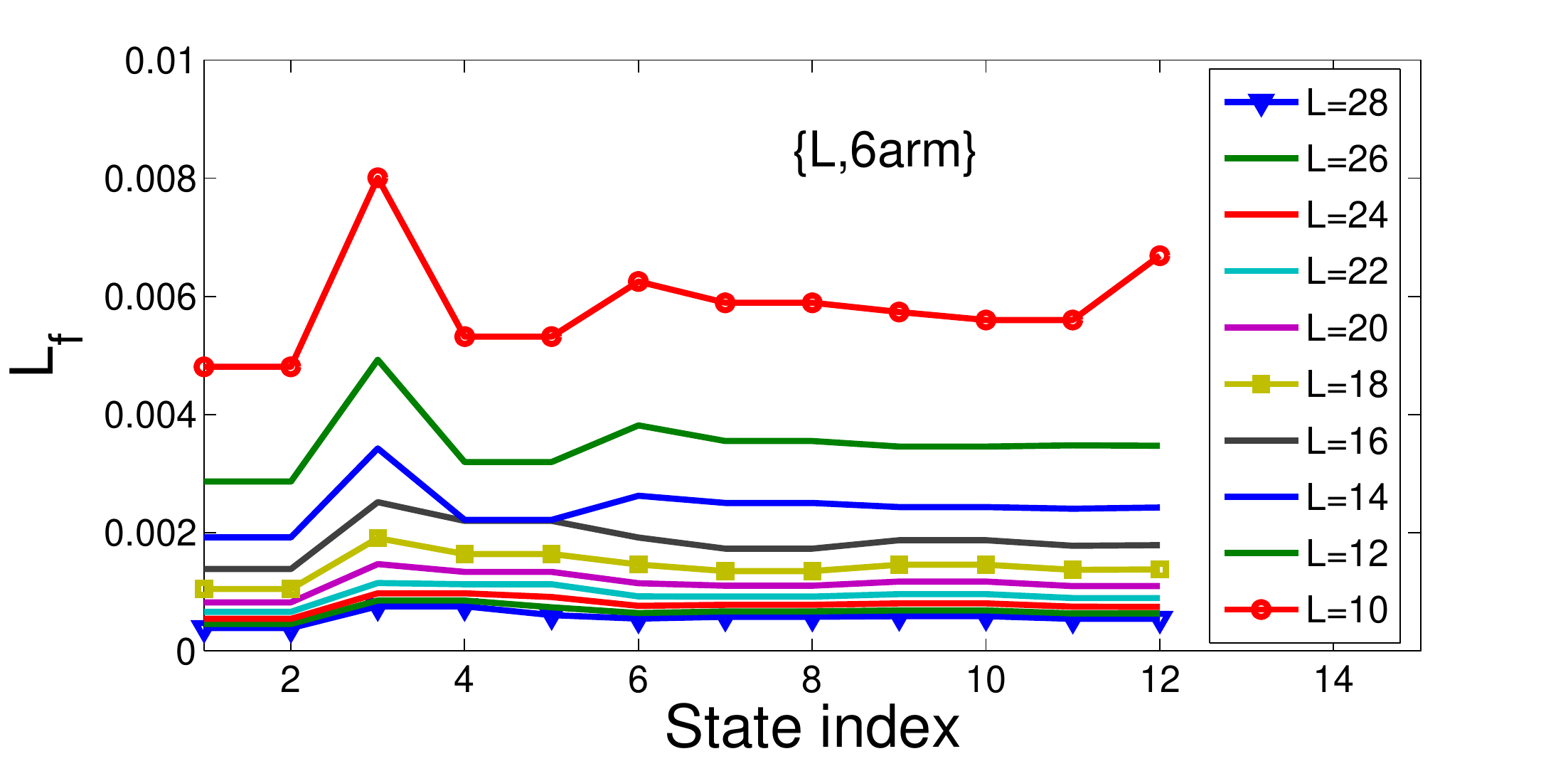}}
\caption{(Color online) Localization factor as a function of conduction state index for a selection of distances between neighboring holes. We compare two fixed hole geometries, namely the \{L,5zz\} (top) and \{L,6arm\} (bottom) GALs. These figures illustrate the physical difference between localization of states in GALs with pure zigzag and armchair edges through the variation of $L$ and thereby the distance between the holes. Localization of states in GALs with zigzag edges is due to edge state localization as opposed to GALs with armchair edges, where localization is a result of the confinement of the electrons.}
\label{fig:LocalizationLdependence}
\end{figure}
A numerical example is given in Fig.~\ref{fig:LocalizationLdependence}.
For the system with $L=10$ and zigzag edges we see that the first conduction states corresponding to the flat minibands are more localized than the following bands. For the armchair edge the localization parameter shows a weaker dependence on the state index. We conclude that the flat minibands of the zigzag edge are more localized than the corresponding states of an antidot lattice with armchair edges. Figure \ref{fig:LocalizationLdependence} furthermore illustrates that this conclusion is independent of the values considered here for the unit cell dimension $L$. The localization factor for the armchair edge depends more strongly on the dimension of the unit cell, that is, the hole-hole distance, and is in general an order of magnitude lower compared to the zigzag edge. Therefore, the band gaps of GALs with armchair edges are determined by the confinement, as opposed to the case of zigzag edges where it is governed by edge state formation. This is the reason why the band gap scales differently depending on the edge type of the hole (Fig.~\ref{fig:Egscaling}). For very small holes with zigzag edges we find that the almost dispersionless GAL minibands are positioned further into the band structure, see the \{L,3zz\} curve in Fig.~\ref{fig:Egscaling}. However, as the length of the edge is increased the zigzag edge-state energies are located directly at the band gap. It is therefore not to be expected that a larger hole with certainty results in a larger band gap. Even though this happens for holes with armchair edges, introduction of zigzag regions may suppress the band gap, which is important for electronic and optical applications of antidot lattices.

We conclude that GALs with armchair edge geometry have a larger band gap as compared to both zigzag edge geometries and the predicted scaling\cite{pedersen_graphene_2008}. Furthermore, the hexagonal antidots with armchair edges show a systematic scaling of the band gap with hole size making this system preferable for electronic applications.

% ********************************************************************************************************
%\newpage
\section{Heat transport\label{Phonon}}
\begin{figure}[htbp] % Fine now (Change to one that goes beoynd M=6,\{10,7zz\} a possibility).
\centering
%{\includegraphics[width=0.3\paperwidth]{FiguresInitial/ThermalConductancePhononHolesPristine.png}}{\includegraphics[width=0.4\paperwidth]{FiguresInitial/ThermalConductancePhononConvV2.png}}%NoV2 0.25%NoV2 0.3
{\includegraphics[width=0.4\paperwidth]{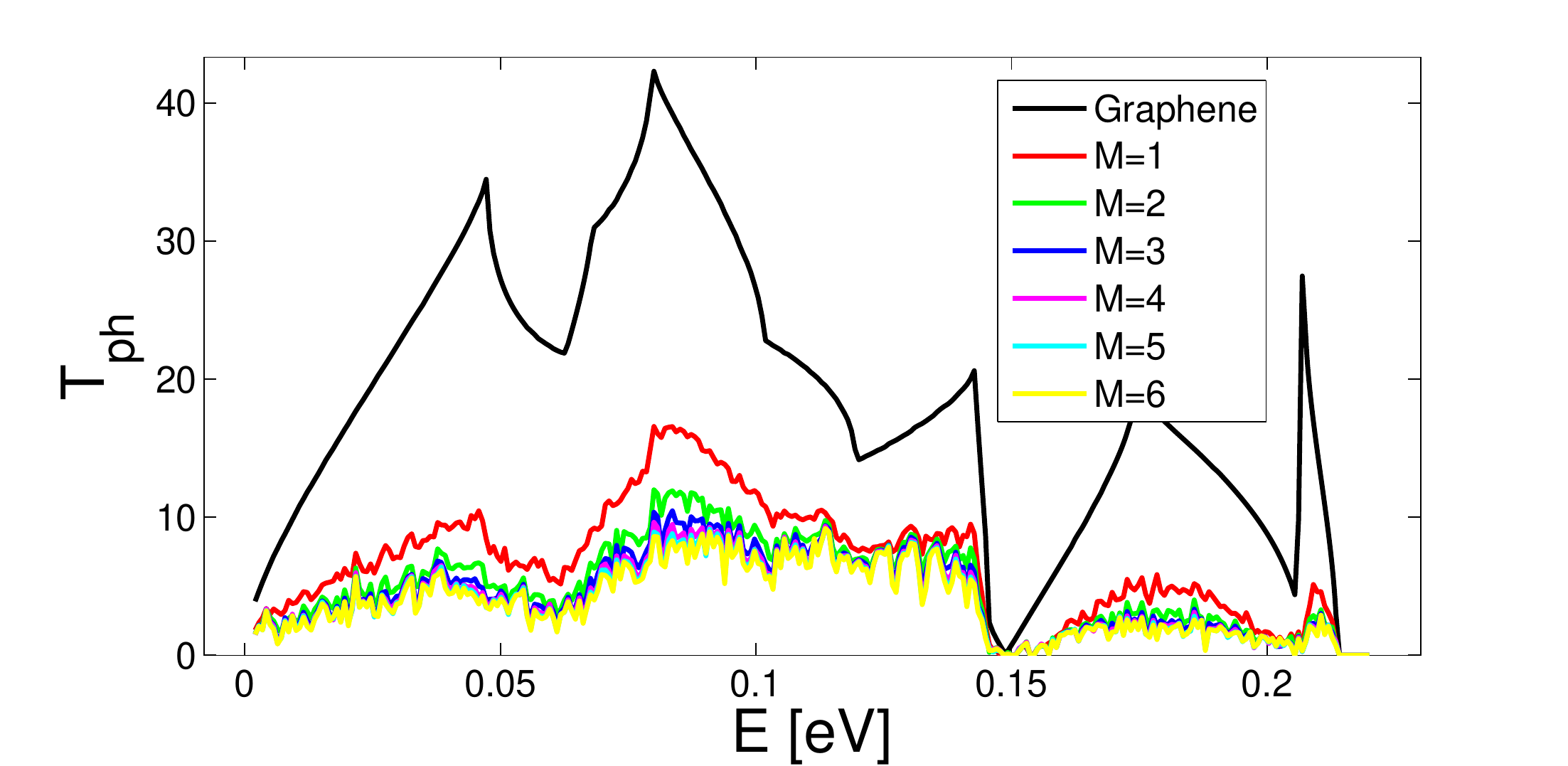}}
{\includegraphics[width=0.19\paperwidth]{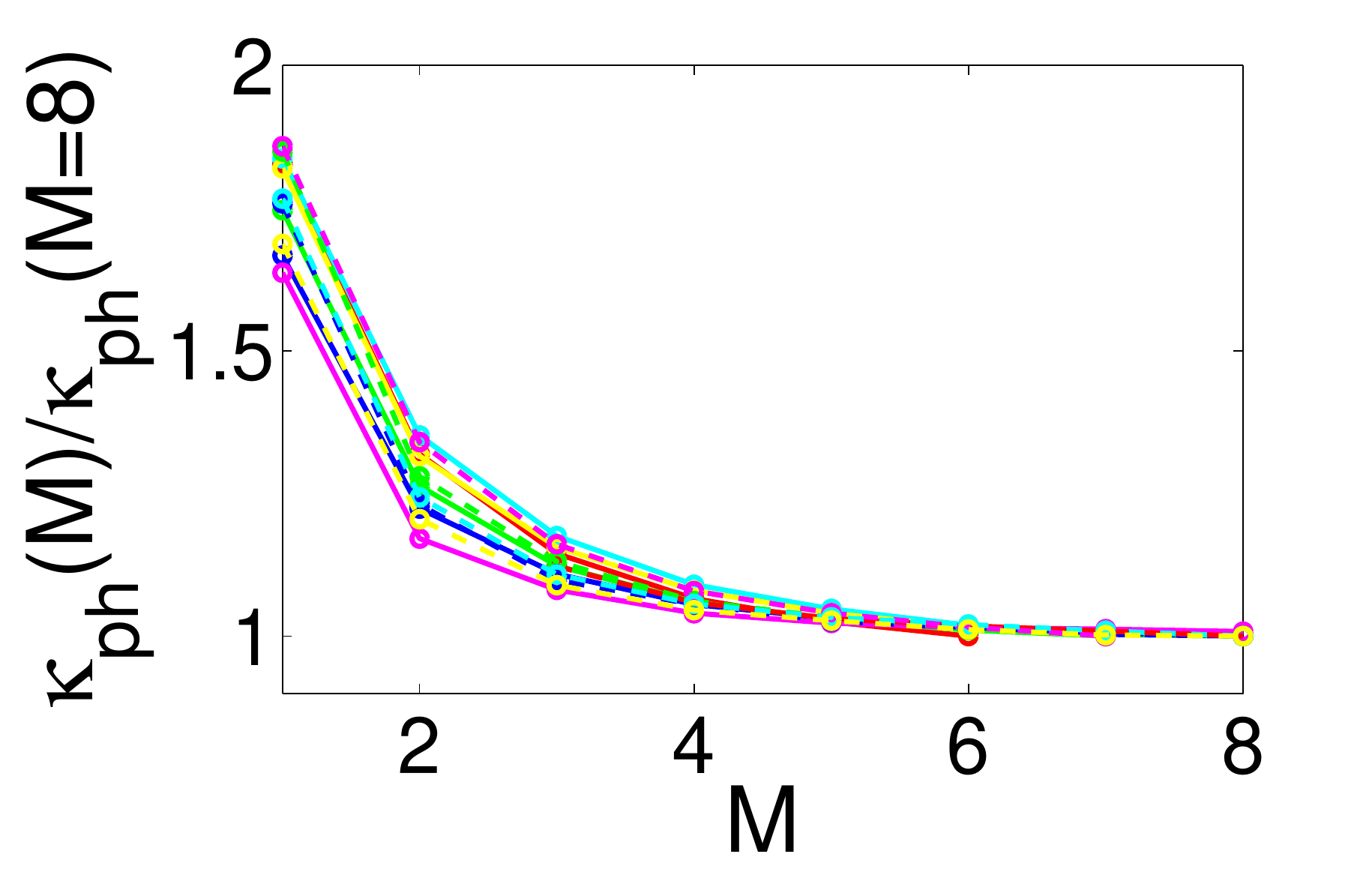}}%3b
{\includegraphics[width=0.19\paperwidth]{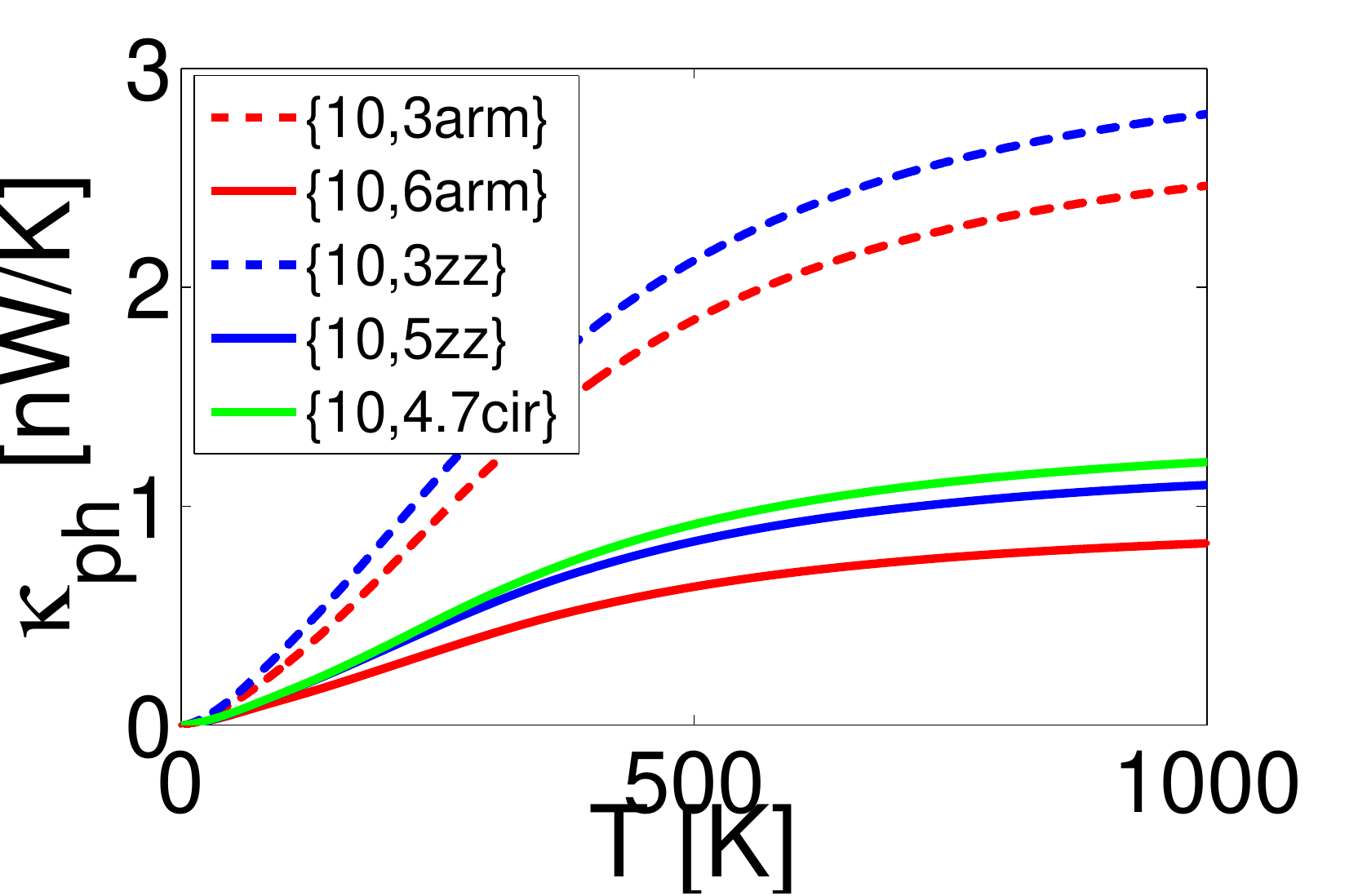}}
%{\includegraphics[width=0.19\paperwidth]{FiguresInitial/TransmissionPhononsNarrow2}}%0.3ThermalConductancePhononConvV2Narrow
\caption{(Color online) Top: Convergence with length of phonon transmission at 300\,K for a \{10,3arm\} antidot lattice.
Bottom left: Convergence of the thermal conductance from phonons with length, normalized by its value at $M$=8. Bottom right: Comparison of the thermal conductance as a function of temperature for the selected GALs shown in Fig.~\ref{fig:Relaxations}.}
\label{fig:ThermalTransmissionConv}
\end{figure}

We now turn to the thermal transport properties of finite GALs. In Fig.~\ref{fig:ThermalTransmissionConv} we show the phonon transmission as a function of $M$, the number of repeated unit cells.
%The result is given in Fig.~\ref{fig:ThermalTransmissionConv}, normalized by the final obtained value.
To quantify the convergence with length we have calculated the thermal conductance at 300\,K for the antidot lattices of Fig.~\ref{fig:Relaxations} at different lengths. This shows that the thermal properties converge at a length scale similar to that of the electrons so also the phonons behaves 'bulk-like' after six to seven unit cell repetitions. In all simulations presented hereafter we use 8-10 unit cells.

The thermal conductance due to phonons in pristine graphene at 300\,K should be compared to a measured thermal conductivity\cite{balandin_superior_2008} of $\sigma_{ph}^{\rm{exp}}\approx 4.5-5.5 \times 10^3 \, \textrm{W}/(\textrm{m K})$. Our result compares well to other theoretical calculations, where it similarly was reported\cite{jiang_thermal_2009} that the reduced ballistic thermal conductance, which we find to be $\kappa_{ph}^{\rm{pri}}/(W_0 h) \approx 4.27 \times 10^{9}\,\textrm{W}/(\textrm{m}^2\textrm{K})$, is much larger than the experimentally extracted partially diffusive result $\sigma_{ph}^{\rm{exp}}/(L^{\rm{exp}}) \approx 0.39-0.48\times 10^{9}\, \textrm{W}/(\textrm{m}^2\textrm{K})$.
Here $h=3.35\,$\AA \, and $L^{\rm{exp}}\approx 11.5\,\mu$m are the graphite interlayer distance and traveled distance by the phonons in the experiment by Balandin \textit{et al.}\cite{balandin_superior_2008}, respectively, and $W_0$ is the computational unit cell width. The main difference here can probably be attributed to isotopes, electron-phonon scattering and especially anharmonicity being important for long devices.
%The thermal conductance from phonons in pristine graphene at 300K is found to be $\kappa_{ph}^{pri}\approx 6.1\textrm{nW}/\textrm{K}$ for the given width of the unitcell $W_0=??$. This should %be compared to a measured thermal conductivity \cite{balandin_superior_2008} of $\sigma_{ph}^{exp}\approx 4.5-5.5 \times 10^3 \textrm{W}/(\textrm{m K})$.

In analogy with the electronic transmission we have calculated an average reduction factor for the phonon transmission.
The transmission of the lowest acoustic and especially the highest optical modes is in general reduced more than the remaining of the phonon spectrum. %... Why?\\
Similarly to the electron case, the average reduction factor decreases linearly with the hole width for the considered systems. The average phonon transmission reduction factor is found to be of the same order of magnitude as compared to the electron transmission. Once again only a minor part of the transmission reduction can be ascribed to the reduction in effective width of the conducting plane due to the perforation. There is a tendency that for small hole sizes the phonons are scattered more than the electrons by the nanoperforation. Furthermore, the electronic reduction factor can be much larger at a specific chemical potential for small holes. For large hole dimensions both the electrons and phonons are scattered to an extent where the transmission is reduced by more than 80\% on average for the systems considered. For the largest holes up to 36\% of the atoms have been removed from the pristine graphene plane.

\begin{figure}[htbp]
\centering
%{\includegraphics[width=0.4\paperwidth]{FiguresInitial/TemperatureThermalConducReorder.png}}
%{\includegraphics[width=0.4\paperwidth]{FiguresInitial/TemperatureThermalConducReorder.png}}%{\includegraphics[width=0.3\paperwidth]{FiguresInitial/TemperatureThermalConducNormReorder.png}}
{\includegraphics[width=0.4\paperwidth]{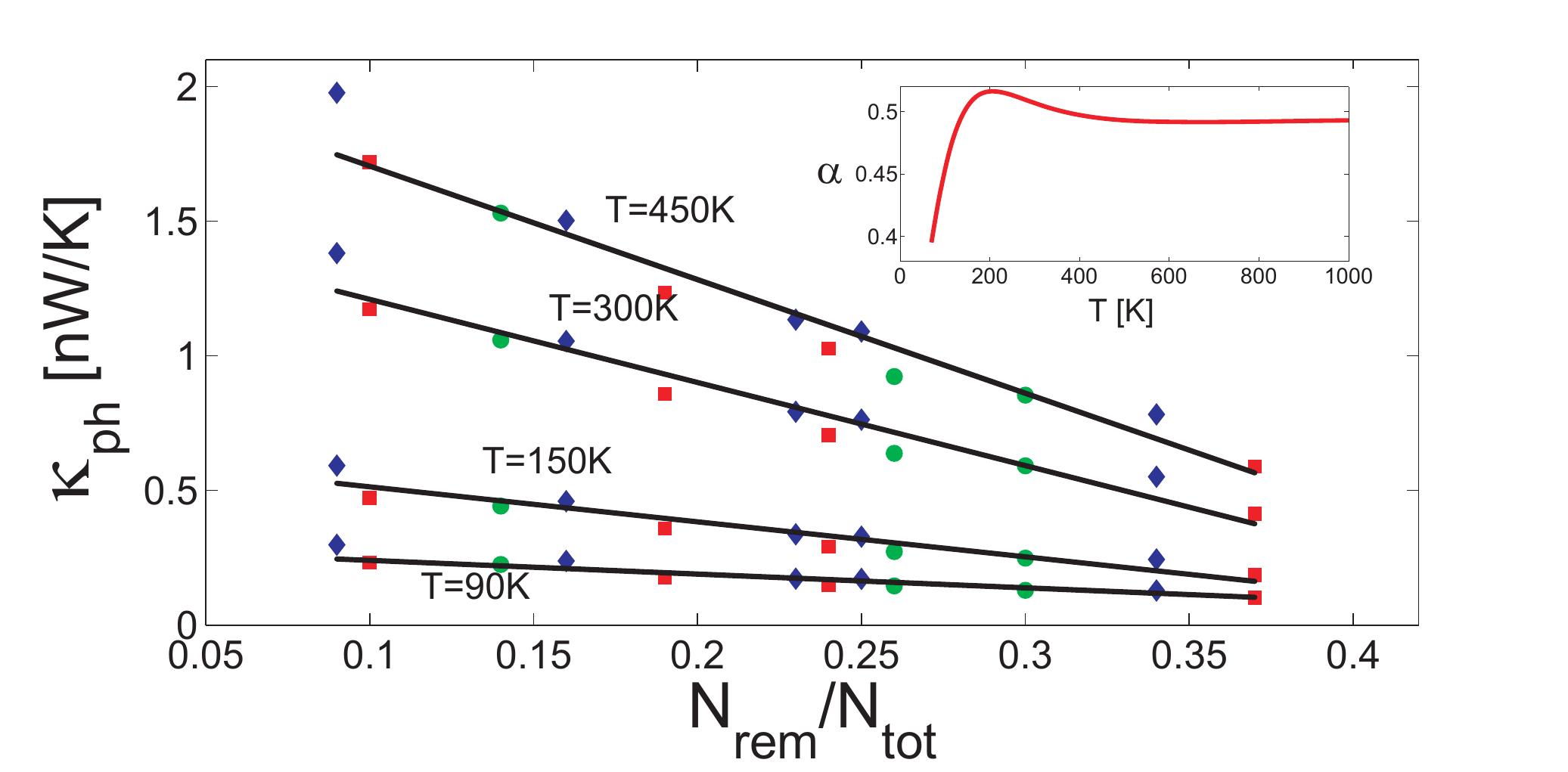}}
%{\includegraphics[width=0.4\paperwidth]{TemperatureThermalConductanceHoleScalingReorderFitV2}} % 0.3
%{\includegraphics[width=0.3\paperwidth]{FiguresInitial/TemperatureThermalConductanceHoleScalingReorderV2.png}}
\caption{(Color online) Thermal conductance from phonons as a function of hole dimension. The red squares, blue diamonds and green circles label holes with armchair, zigzag and circular/mixed edges respectively. Four different temperatures are plotted for each system. From top to bottom the thermal conductance is found at a temperature [450,300,150,90]\,K. The thermal conductance at these four temperatures is for pristine graphene found to be [8.5,6.1,2.6,1.2]$\frac{\textrm{nW}}{\textrm{K}}$. Inset: fitted dimensionless parameter $\alpha$ describing the scaling with hole size of the thermal conductance.}
\label{fig:PhononConductanceTemp}
\end{figure}

In Fig.~\ref{fig:PhononConductanceTemp} the temperature and hole size dependence of the phonon thermal conductance is given for our selection of systems with varying hole size and shape.
Fig.~\ref{fig:PhononConductanceTemp} illustrates how the thermal conductance decreases almost linearly with the hole size for typical perforation removal ratios (larger than $5\%$ perforation). Furthermore, the graph shows that the thermal conductance has a tendency to be slightly larger for holes with zigzag edges (shown as diamonds in Fig.~\ref{fig:PhononConductanceTemp}). A similar behavior has been found for graphene nanoribbons with zigzag edges\cite{hu_thermal_2009}. However, compared to the electronic case the thermal transport features are less sensitive to the exact shape and edge of the holes.

For the purpose of making it easy to compare our result with other calculations and experiments we give an empirical expression for the thermal conductance. In the regime where the thermal conductance is linear in the hole dimension one can parameterize the thermal conductance as
\begin{eqnarray}
%\frac{\kappa_{ph}}{W h} \approx -\alpha(T) \frac{N_{rem}}{N_{tot}}+\beta \frac{\kappa_{ph}^{pri}(T)}{W h} + \gamma_0
\kappa_{ph}\approx \left(-\alpha(T) \frac{N_{rem}}{N_{tot}}+\beta\right) \kappa_{ph}^{\rm pri}(T)\,.
\end{eqnarray}
From this approximation we fit the 'linear regime offset' $\beta \approx 0.25$ and the dimensionless parameter $\alpha(T)$, given in the inset of Fig.~\ref{fig:PhononConductanceTemp}.
The lines in Fig.~\ref{fig:PhononConductanceTemp} are illustrating this parametrization.

\begin{figure}[htbp]% plot for a couple of lattices at 300K
\centering
%{\includegraphics[width=0.195\paperwidth]{TemperatureThermalConducElecc}}
%{\includegraphics[width=0.195\paperwidth]{TemperatureThermalConducEleczz9c}}% 0.25
{\includegraphics[width=0.4\paperwidth]{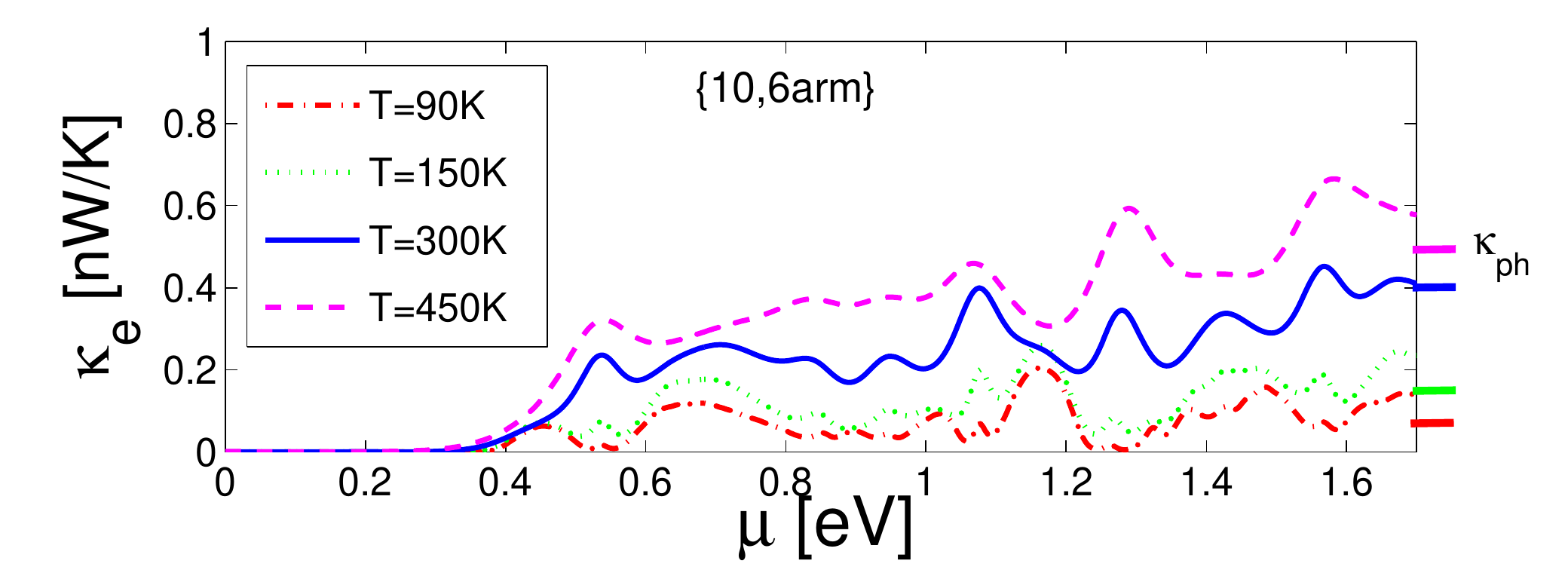}}
{\includegraphics[width=0.4\paperwidth]{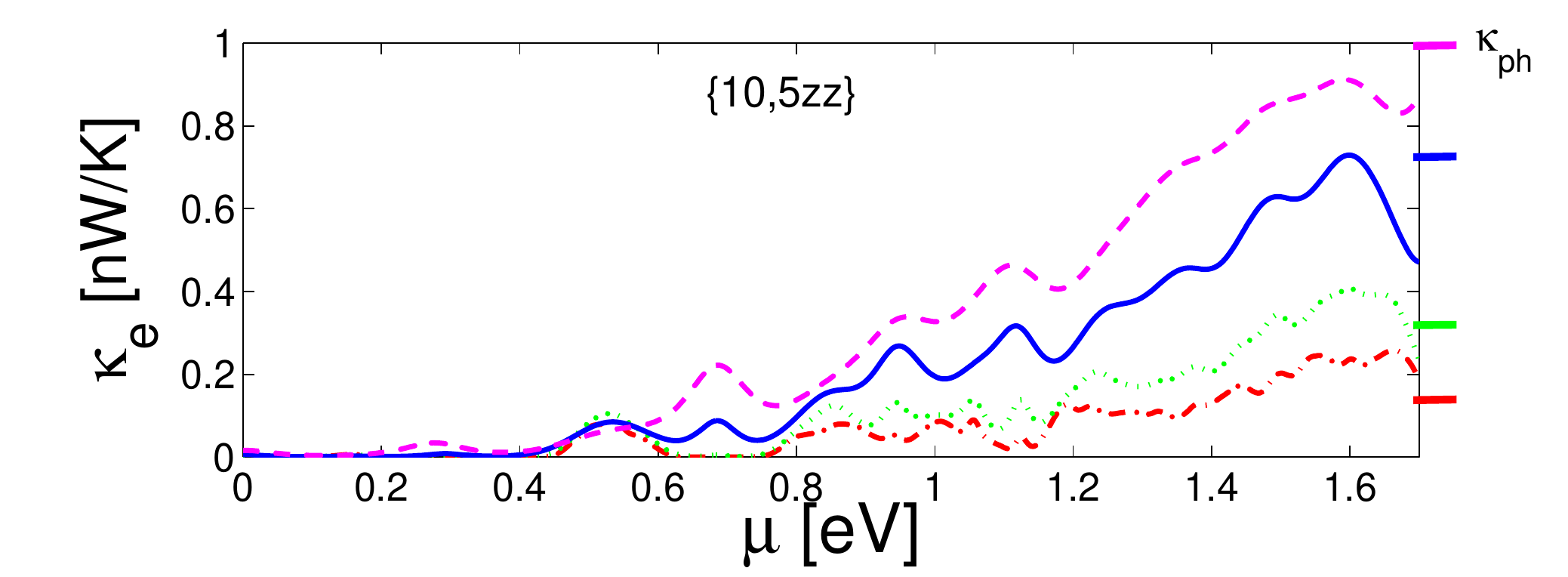}}% 0.25
\caption{(Color online) Electronic contribution to the thermal conduction. Top: \{10,6arm\} antidot lattice. Bottom: \{10,5zz\} antidot lattice. From top to bottom the curves are found at a temperature of [450,300,150,90]\,K. The corresponding phonon thermal conductance has been marked to the right of the plot for comparison.}
\label{fig:TempThermalConducElec}
\end{figure}

In Fig.~\ref{fig:TempThermalConducElec} we have illustrated the electronic contribution to the thermal conductance at four different temperatures.
% (pink,blue,green and red respectively).
Due to the vanishing electronic density of states around zero chemical potential the thermal conductance of GALs is dominated by phonons as is also the case of pristine graphene. However, the electronic contribution can dominate - even at room temperature - when a large gate bias is applied.

%**********************************************************************************************
\section{Thermoelectric figure of merit\label{ZT}}
\begin{figure}[htbp] % plot for a couple of lattices at 300K
\centering
%{\includegraphics[width=0.3\paperwidth]{FiguresInitial/TemperatureSeebeck.png}}
%{\includegraphics[width=0.4\paperwidth]{FiguresInitial/10arm/TemperatureSeebeckV2.png}}\\%TemperatureSeebeckV2Zoom.png}}
%{\includegraphics[width=0.3\paperwidth]{FiguresInitial/10arm/TemperaturePowerFactorV2.png}}\\
%{\includegraphics[width=0.4\paperwidth]{FiguresInitial/9zz/TemperatureSeebeckV2.png}}\\
%{\includegraphics[width=0.3\paperwidth]{FiguresInitial/9zz/TemperaturePowerFactorV2.png}}
{\includegraphics[width=0.4\paperwidth]{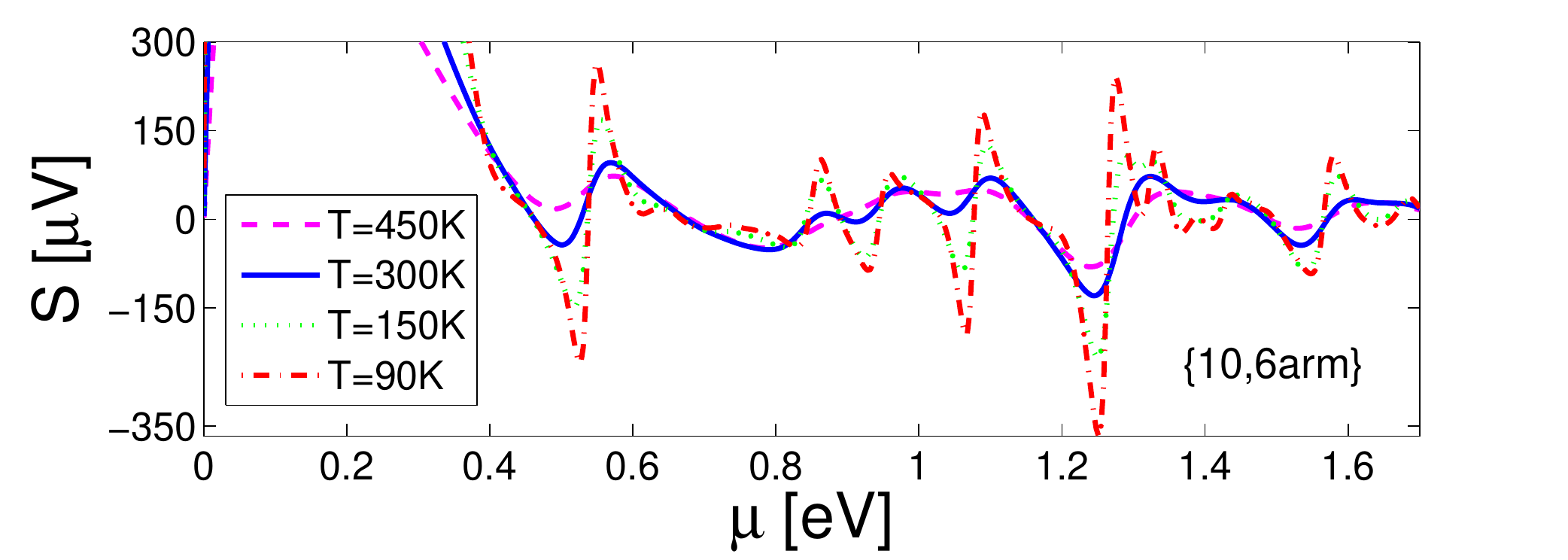}}
{\includegraphics[width=0.4\paperwidth]{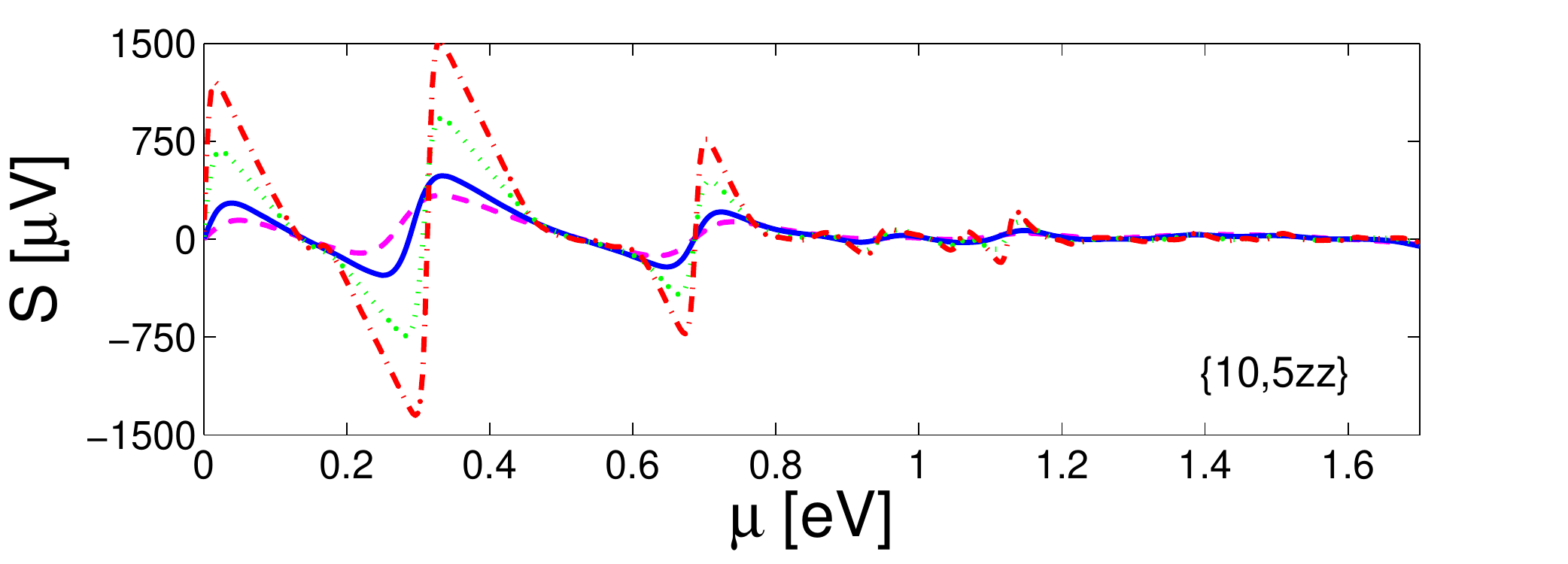}}
\caption{(Color online) Seebeck coefficient at the four different temperatures [450,300,150,90]\,K for the \{10,6arm\} (top) and \{10,5zz\} (bottom) GAL. Notice the different scale on the Seebeck coefficient. The main Seebeck peaks of the GAL with zigzag edges occur at the energies of the low transmitting localized states, whereas the main contribution to the thermoelectric power of the GAL with armchair edges is positioned at energies further into the band structure.}
\label{fig:TempSeebeck}% and power factor
\end{figure}

Next we report the thermoelectric properties of the considered GALs. In Fig.~\ref{fig:TempSeebeck} we compare the Seebeck coefficient for a GAL with armchair edges (top) and zigzag edges (bottom).
% Again the four temperatures [450,300,150,90]K are colored pink,blue,green and red respectively.
The Seebeck coefficient displays peak values\cite{Note3} of the order of 0.1-1.5\,mV/K, which is similar to what has been obtained for other carbon based nanosystems and molecular contacts\cite{saha_multiterminal_2011,ouyang_theoretical_2009,bao_thermoelectric_2010}.

\begin{figure}[h!tbp] %plot for a couple of lattices at 300K
\centering
%{\includegraphics[width=0.35\paperwidth]{TemperatureZTV2}}\\
%{\includegraphics[width=0.35\paperwidth]{TemperatureZTelNew}}\\%elV2.png}}
%{\includegraphics[width=0.35\paperwidth]{TemperatureZTzz}}\\
%{\includegraphics[width=0.35\paperwidth]{TemperatureZTelzzNew}}
%%{\includegraphics[width=0.35\paperwidth]{ZTcombob}}
{\includegraphics[width=0.35\paperwidth]{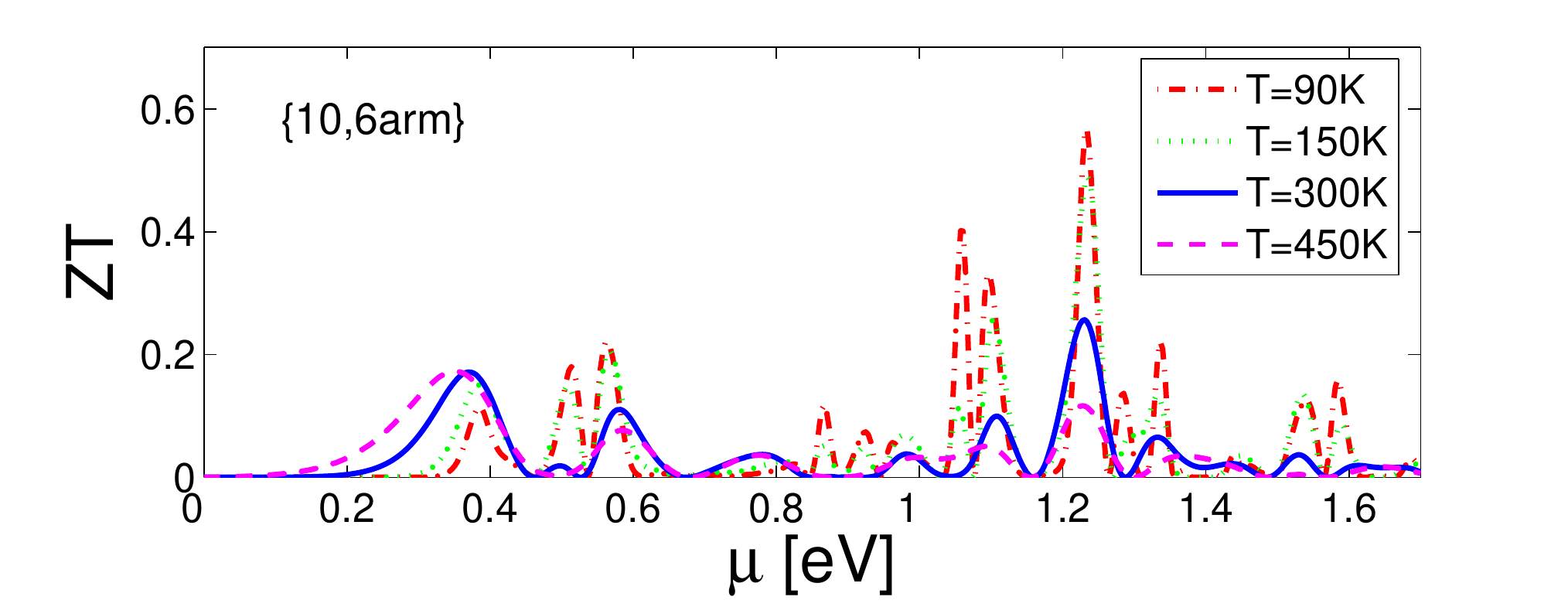}}
{\includegraphics[width=0.35\paperwidth]{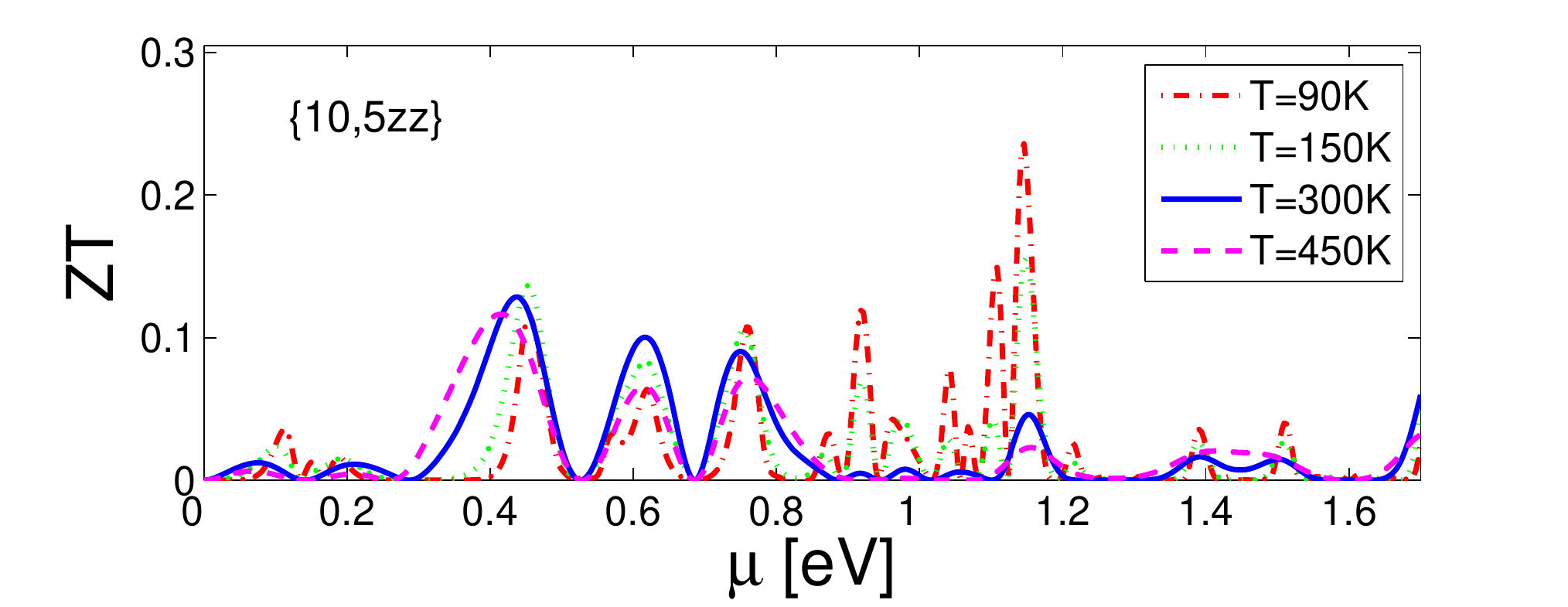}}
\caption{(Color online) $ZT$ for \{10,6arm\} and \{10,5zz\} lattices at the four temperatures [450,300,150,90]\,K. At low temperature the pure electronic figure of merit, $ZT_{el}$, can be very large due to a vanishing thermal conductance from electrons and sharp features in the transmission spectrum. At low temperature many sharp transmission features also becomes visible in the actual $ZT$.}
\label{fig:TempZtZtel}
\end{figure}

\begin{figure}[h!tbp]
\centering
{\includegraphics[width=0.4\paperwidth]{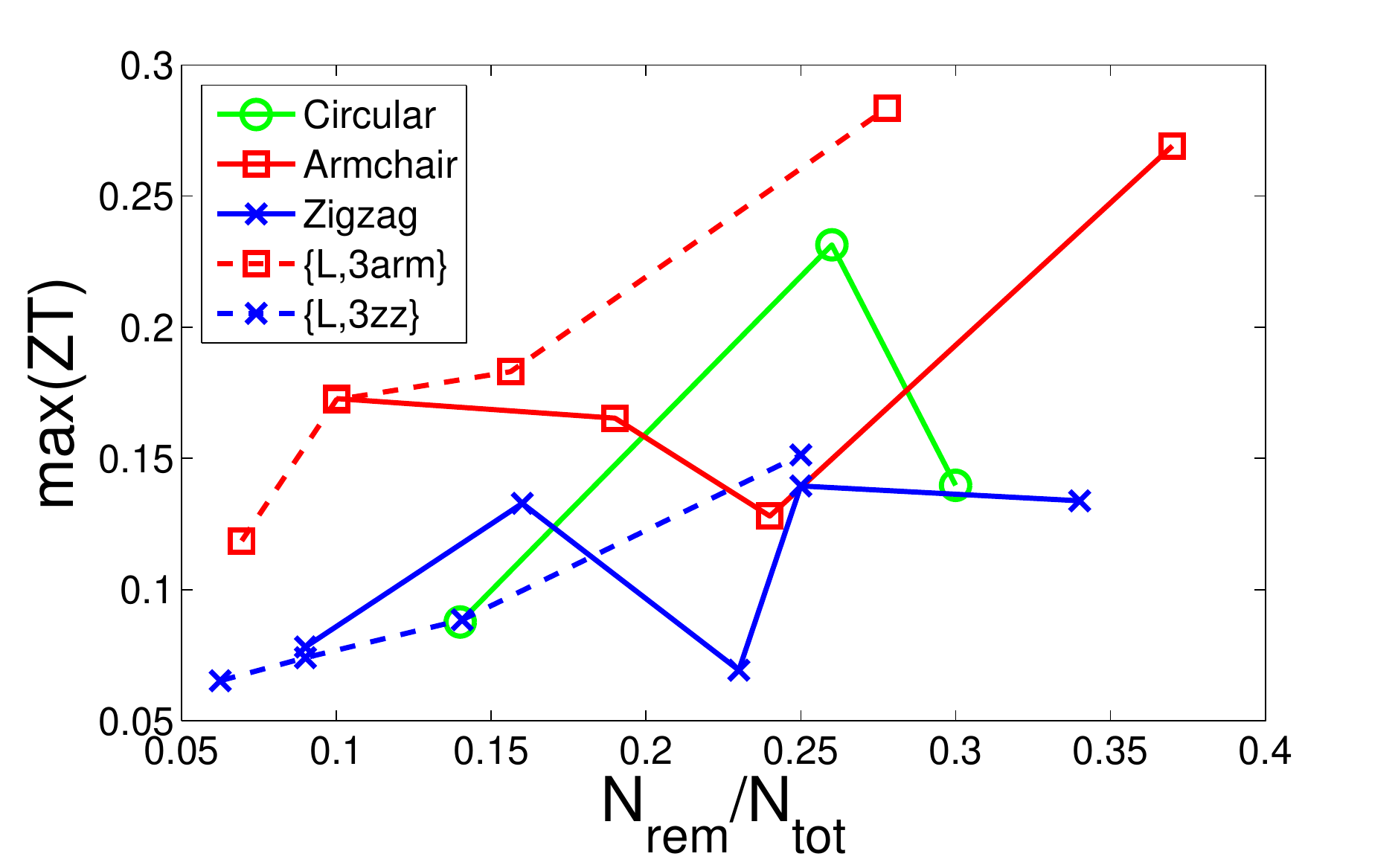}}
\caption{(Color online) ZT dependence on the ratio of removed atoms in the nanoperforation at $T=300$\,K. Systems included in the figure are \{10,$S$zz\} with $S$=3,4,4.5,5,5.5, \{10,$S$arm\} with $S$=3,4.5,5,6 and \{10,$S$cir\} with $S$=3.5,4.7,5 (full lines) and furthermore two set of systems with a fixed hole \{$L$,3arm\} and \{$L$,3zz\} with $L$=6,8,10,12 (dotted lines).}
\label{fig:ZTscaling}
\end{figure}

For bulk material the thermoelectric figure of merit is defined in terms of the electrical and thermal conductivities, $\sigma_e,\sigma_t$, as $ZT=T \sigma S^2/\sigma_t$. For the ballistic graphene systems we can write it in terms of their respective conductances by introducing a width, effective length and thickness, $ZT= T G_e S^2/(\kappa_e+\kappa_{ph})$.
Maximal thermoelectric figure of merit, $ZT$, is obtained after length convergence due to the increased band gap and decreased thermal conductance. The obtained $ZT$ shown in Fig.~\ref{fig:TempZtZtel}, as a function of chemical potential has a number of peaks corresponding to a large variation of the transmission with energy. The Seebeck coefficient is a measure of these changes and their robustness to temperature smoothening. One important feature is that the high peaks in the Seebeck coefficient for the  \{10,5zz\} lattice mainly occur at very low energy, where the transmission is low, whereas for the \{10,6arm\} lattice the dominating peaks occur at higher chemical potential. Therefore, the peak $ZT$ is higher for the \{10,6arm\} as a result of the higher electronic conductance at peak position as illustrated in Fig.~\ref{fig:TempZtZtel}.
% MB: Need to compare to something!!
%The reason for the increased thermoelectric figure of merit in GALs is found to be a significant increase in the Seebeck %coefficient together with a relative reduction of the electronic to thermal conductance. 
The Seebeck coefficient is highly sensitive to the variations in the electronic transmission resulting from different hole edges, sizes and so forth. In Fig.~\ref{fig:ZTscaling} we collect the maximum $ZT$ values we have found for a selection of GALs.
It seems possible to obtain larger $ZT$ from GALs based on hexagonal holes with armchair edges. This is a result of the additional splitting into minibands for zigzag edges. The reason for this additional splitting is, as mentioned in \ref{Sect:Localized}, the formation of edge states at zigzag edges. As a consequence the Seebeck coefficient can be larger for zigzag edges. However, the power factor is significantly lower due to the lower transmission from the isolated energy levels with low dispersion. There is also a weak trend that the hole dimension compared to the system size should be maximized. By increasing the hole dimension we actually reduce the electronic figure of merit, defined as
\begin{eqnarray}
ZT_{el} = \frac{S^2 G_e T}{\kappa_e} = \frac{\kappa_{ph}+\kappa_e}{\kappa_e} ZT\label{eqn:ElecZT}\,,
\end{eqnarray}
but obtain a larger fraction of it due to a reduced phonon conductance.
Higher $ZT$ could possibly be obtained by increasing the hole dimension even further, but these systems will be very challenging to fabricate.

The electrons-only result, $ZT_{el}$ ($\kappa_{ph}=0$) describes an upper bound of the figure of merit. However, we find it to be somewhat artificial, due to the fact that the phonon contribution to the thermal conductance shifts the position of the peaks and $ZT_{el}$ posses a (in principle unbound) peak every time the electronic thermal conductance is zero (Eq. (\ref{eqn:ElecZT})). Especially in the presence of gaps in the electronic band structure, the computation of $ZT_{el}$ can be numerically challenging. However, evaluating the  $ZT_{el}$ expression at the true peak position can give an estimate of the gain by a further reduction of the phonon conductance. For the \{10,6arm\} GAL the first peak ($\mu=0.37$\,eV) and the highest peak ($\mu=1.23$\,eV) have a $ZT=0.17$ and $0.26$ with corresponding $ZT_{el}=4.78$ and 0.77, a factor of 28 and 3 larger than the true $ZT$, respectively. For the \{10,5zz\} lattice we have the first peak value $ZT=0.13$ with corresponding $ZT_{el}=5.77$ a factor of 44 larger at the same energy. At high chemical potential the main limitation is the electronic structure and not a further reduction of the phonon heat conductance. On the contrary one could obtain a significant $ZT$ enhancement at low chemical potential by further reducing the thermal conductance. Isotope scattering, anharmonic interactions, electron-phonon interactions and graphene-substrate interactions could all contribute to a reduction of the phonon thermal conductance.
\begin{figure}[h!tbp]
\centering
{\includegraphics[width=0.19\paperwidth]{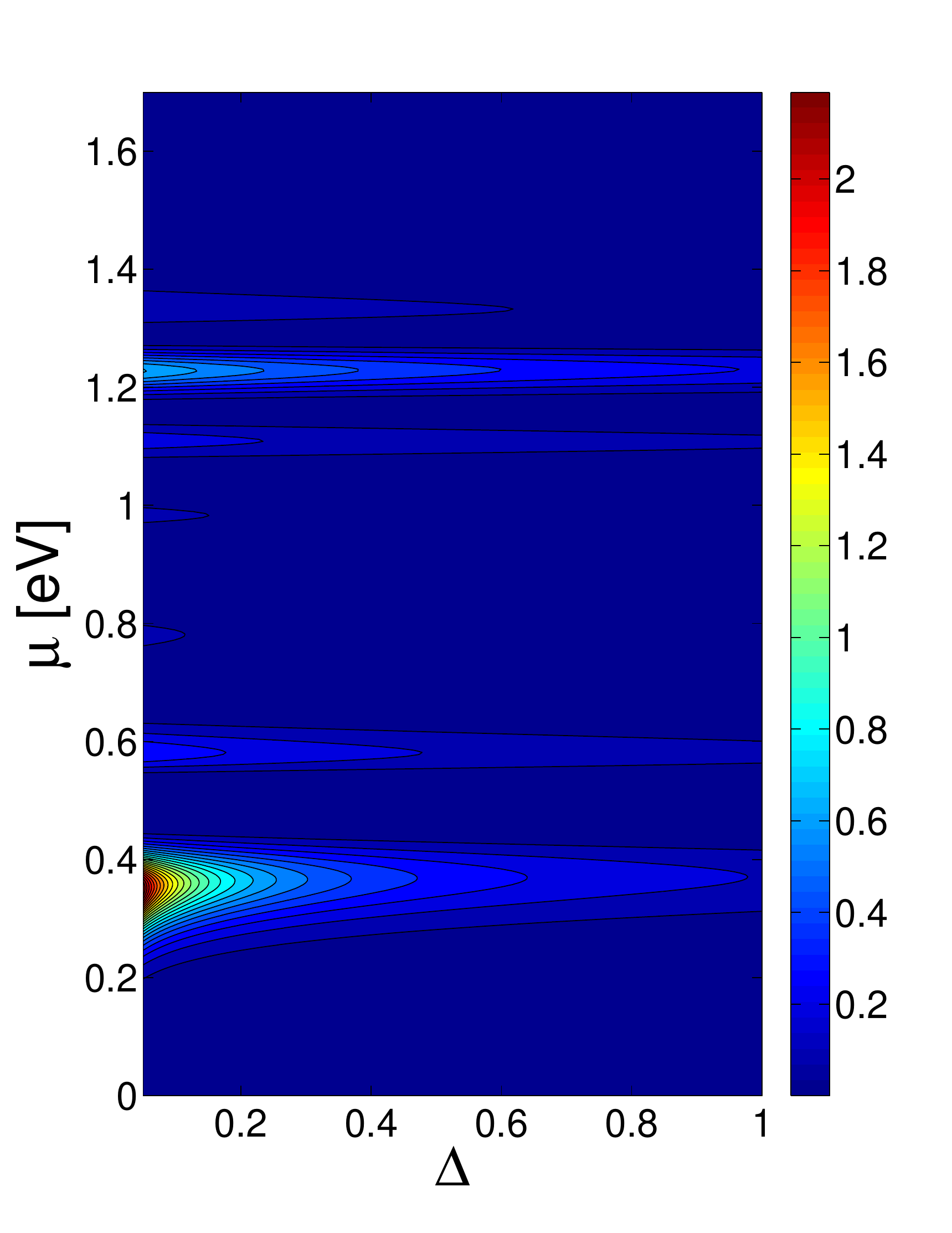}}
{\includegraphics[width=0.19\paperwidth]{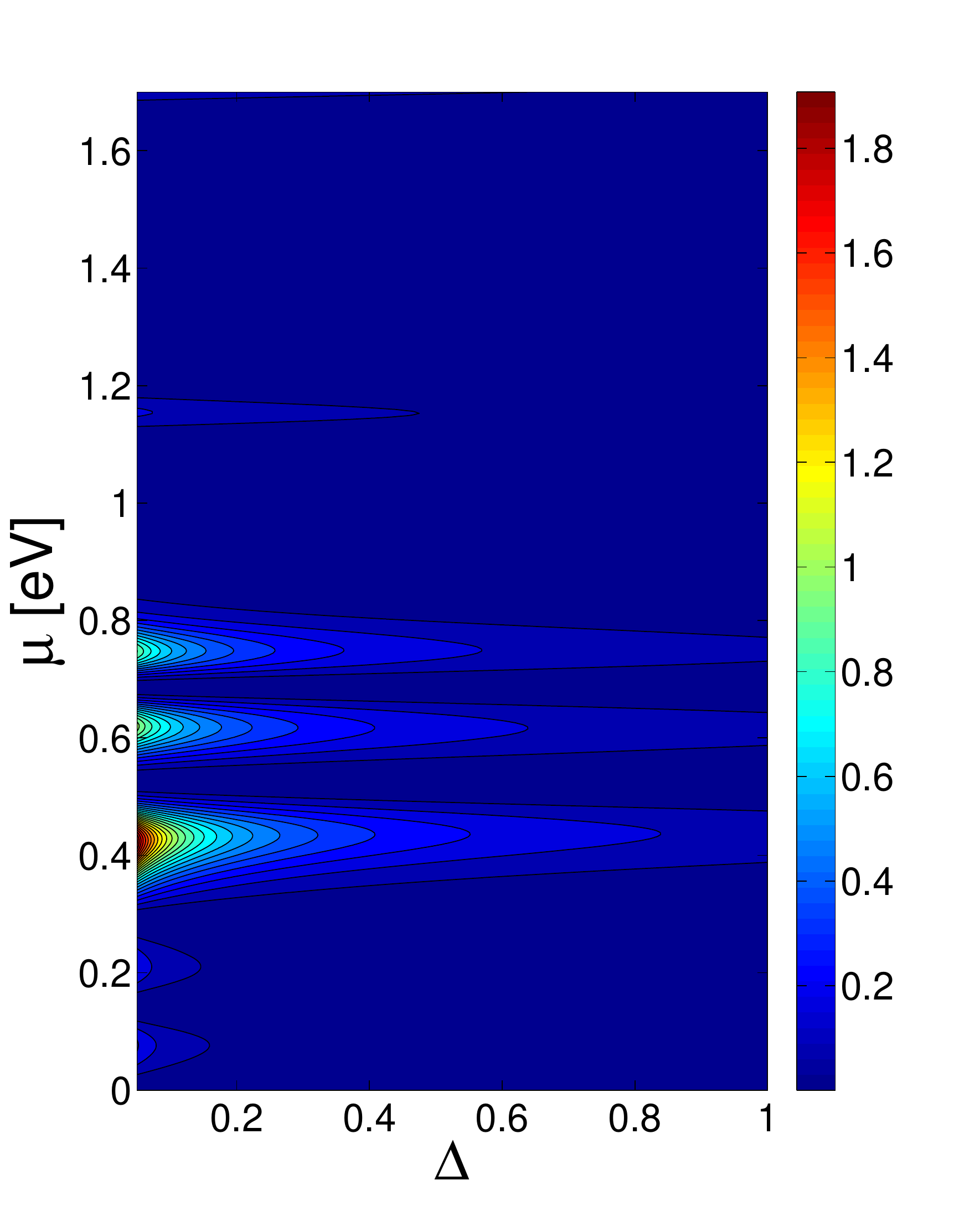}}
{\includegraphics[width=0.35\paperwidth]{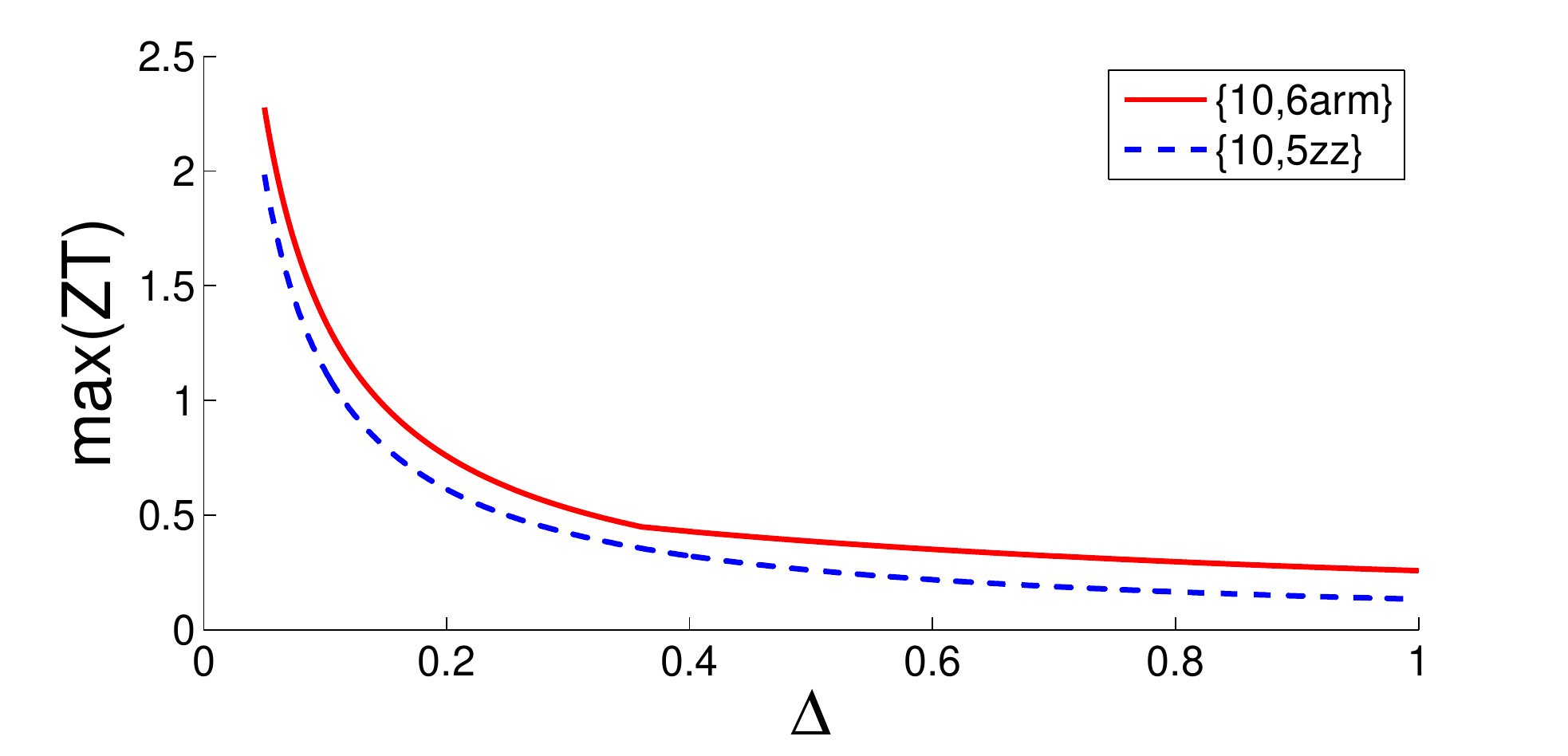}}
\caption{(Color online) ZT variation with a decrease of the phonon thermal conductance at $T=300$\,K. $\Delta$ describes the fraction of the original thermal conductance used in the calculation. Top: $ZT$ as a function of the chemical potential and the phonon thermal conductance for the \{10,6arm\} (left) and \{10,5zz\} (right) GAL. Bottom: Peak $ZT$ as a function of the variation of the phonon thermal conductance.}\label{fig:ZTnewKappa}
\end{figure}
In Fig . \ref{fig:ZTnewKappa} we illustrate the effect of a reduction of the phonon thermal conductance. The parameter $\Delta$ gives the fraction of the original phonon thermal conductance kept in the calculation. For the \{10,6arm\} GAL the first peak ($\mu \approx 0.37$\,eV) increases more rapidly than the high energy peaks. The peak position crossover happens around $\Delta\approx$0.35. When the phonon thermal conductance dominates, the figure of merit variation goes as $ZT/\Delta$, clearly present in the low energy $ZT$ variation (see Fig.~\ref{fig:ZTnewKappa}, bottom) even when the phonon thermal conductance is reduced below 5\% of its original value.

We propose one of two routes to obtaining higher $ZT$. Either one could find a way to reduce the thermal conductance without affecting the electrons. Surface decoration might be a promising way to obtain this. Another route could be to improve the electronic properties of GAL's, e.g. by combining this system with other nanostructured devices. This could increase the peak $ZT$ obtained at high chemical potential. We note that one-dimensional nanosystems may display a large Seebeck coefficient leading to larger $ZT$. However, on the other hand the two-dimensional GAL structure may have an important advantage.
In a 2D GAL the system geometry does not pose inherent limits for the converted power, as it does for a system of parallel quasi-one-dimensional systems, such as quantum wires or graphene nanoribbons: the power can be increased by simply making the GAL-system broader. This provides a strong motivation for further investigations of GALs as thermoelectric devices, perhaps as an integrated element in future graphene nanoelectronics. Despite the high intrinsic thermal conductance of graphene it is noticeable that one can utilize the nanoperforation to obtain $ZT$ exceeding 0.25, a factor 35 enhancement compared to what is found for bulk graphene.
%This might be a reason that the antidot lattices could find application as thermoelectric components integrated in future graphene based nanoelectronics.
%With a 2D material there is no limit to the power one is able to apply to the device. One can simply make the GAL wider in order to increase the power converted, thereby avoiding parallel connections of low dimensional systems.
% One has to be aware that the antidot lattice is truely a nanostructered material. This is important since eventhough lower dimensional nanosystems might posses larger ZT's due to a significant increasement in the Seebeck coefficient one needs to include many such devices in parallel or serial to get good performance in devices. However, with a 2D material as proposed here there is no limit to the power one is able to apply to the device. One can simply make the antidot lattice wider to increase the power converted.
% udvalgt Seebeck, transmissions sammenligning-> plot Tph ved M = 1 og 5 i samme plot og tilsvarende for Te-> relativ reduktion. Plot seebeck ved disse 2 også (mindre dramatisk ændring).
% bandgap for M>1. Thermal conductance of graphene antidot lattices is at zero potential dominated by phonons. However, the electronic contribution can dominate when a large gate bias is applied even at room temperature.

\section{Discussion and conclusion\label{Conclusion}}
% Model: It is essential to model both electronic and phononic transport properties at the same level of approximation.
%We have shown that GALs pave the way for manipulation of both the electronic and thermal transport properties of graphene sheets.
We have theoretically shown that GALs allow the simultaneous manipulation of both electronic and thermal transport properties of graphene sheets. Our calculations have been carried out in the ballistic limit, which gives a reasonable first estimate for short devices whose dimensions are smaller than the various scattering lengths (important scattering mechanisms include the anharmonic phonon-phonon interactions, electron-phonon scattering, and electron-electron scattering). Also spin-polarization may turn important: recent studies have shown that one can have spin-splitting and a magnetic moment in triangular\cite{yang_inducing_2011,liu_band-gap_2009,zheng_effects_2009} antidots with pure zigzag edges. Above all, the most important future task is a systematic study of disorder effects. Our preliminary results suggest that a low degree of disorder can increase $ZT$ due to a decrease of the thermal conductance, whereas a high degree of disorder affects both electrons and phonons so that the decrease in power factor outweights the decrease in thermal conductance. 
%We plan to address this issue in future publications.
%We have based our analysis upon atomistic models of both the electronic structure and phonon transport properties and considered the thermal conductance of both electrons and phonons. We have %neglected high order scattering from the anharmonic phonon-phonon interactions, and also the electron-phonon and electron-electron interactions. This approximation is expected to be good for %device regions of short length. The analysis aims at a first estimation of the relative reduction between the electron and phonon transmission. Furthermore, we did not include effects of %spin-polarization, which was recently shown to introduce spin-splitting and a magnetic moment in triangular\cite{yang_inducing_2011,liu_band-gap_2009,zheng_effects_2009} antidots with pure %zigzag edges.

A key result of our analysis is the convergence of transport properties with length for GALs. The ballistic transport properties converge fast towards that of the infinite antidot lattice.
We have also found that the quantization is an important feature of both electron and phonon transport properties of GALs. This is seen from the fact that the transmissions are reduced far more than what would be expected from an effective width estimation and therefore the exact scattering rate for the different edge types is important. The average transmission reduction factor is found to be on the same order of magnitude for electrons and phonons. In general, the formation of edge states determine the band gap of GALs with pure zigzag edges as opposed to pure armchair edges, where the band gap is determined by the confinement of electrons. Furthermore, the different edge characteristics play an important role in the observed difference in thermoelectric properties. $ZT$ is found to be lower for GALs with zigzag edges due to the additional splitting into minibands for large structures and a corresponding lower power factor. The maximal thermoelectric efficiency $ZT\approx0.3$ has been obtained for GALs with pure armchair edges. Therefore, it is possible to obtain fair thermoelectric properties of graphene-based nanosystems, even despite of lattice distortions which highly affect both the $\pi$-electron determined electronic properties and the sp$_2$-bonding determined thermal conductance, such as the nanoperforations. The main limitation in thermoelectric applications of GALs at high chemical potential is set by the electronic structure because the electronic heat conductance is large at the high energy peak position of $S$ and $ZT$. At low chemical potential we expect that one could benefit from a further reduction of the phononic thermal conductance due to isotope scattering and anharmonic interactions.

\begin{acknowledgments}
TM acknowledge support from FTP Grant No. 274-08-0408. APJ is grateful to the FiDiPro program of Academy of Finland.
\end{acknowledgments}

\end{document}